\documentclass[11pt,a4paper]{article}
\usepackage{color,amsmath,amssymb,graphicx,latexsym,lineno}
\usepackage{threeparttable,multirow,txfonts,booktabs}
\usepackage{accents,slashed,ulem,subfig}
 \usepackage{float}
 \usepackage{flushend}
 \usepackage{balance}

\usepackage{jcappub}

\begin{document}

\title{Exploring Constraints on Axion-like Particles with the Observations for blazar Mrk 501}

\author[a]{Ting Zhou,}
\author[b,1]{Lin-Qing Gao, \note{Corresponding author.}}
\author[c,d]{Xiao-Jun Bi,}
\author[c]{Peng-Fei Yin,}
\author[a,1]{Wenbin Lin}

\affiliation[a]{School of Mathematics and Physics, University of South China,
Hengyang, 421001, China}
\affiliation[b]{School of Physics and Electronic Sciences, Changsha University of Science and Technology, Changsha, 410114, China}
\affiliation[c]{Key Laboratory of Particle Astrophysics, Institute of High Energy Physics,
Chinese Academy of Sciences, Beijing, 100049, China}
\affiliation[d]{School of Physical Sciences, University of Chinese Academy of Sciences, Beijing, 100049, China}
\affiliation[e]{School of Physical Sciences and Technology, Southwest Jiaotong University, Chengdu, 610031, China}

\emailAdd{20222004110237@stu.usc.edu.cn}
\emailAdd{gaolq@csust.edu.cn}
\emailAdd{bixj@ihep.ac.cn}
\emailAdd{yinpf@ihep.ac.cn}
\emailAdd{lwb@usc.edu.cn}

\abstract{Oscillations between axion-like particles (ALPs) and photons in astrophysical magnetic fields can lead to irregularities in the high energy gamma ray spectra of blazars. The magnetic field within the blazar jet plays a crucial role in shaping these effects, with its strength in the emission region being an important parameter determined by multi-wavelength observations. However, the origin of the high energy bump observed in the spectral energy distribution of some blazars is a topic of debate, with both leptonic and hadronic scenarios providing plausible explanations that result in different magnetic field strengths in the emission region. In this study, we investigate the impact of magnetic field configurations on the constraints of ALP parameters. We consider both leptonic and hadronic emission scenarios for the blazar Mrk 501 and derive the corresponding jet magnetic field strengths.  Additionally, we explore two jet magnetic field models: one with a toroidal component and the other with helical and tangled components. By analyzing the spectra of Mrk 501 observed by MAGIC and Fermi-LAT from 2017-06-17 to 2019-07-23, which are well-described by both emission scenarios, we derive constraints on the ALP parameters. Our results demonstrate that both the emission scenario and the magnetic field structure play a significant role in deriving these constraints, with the hadronic model leading to much more stringent limits compared to the leptonic model.}
\arxivnumber{2501.18860}
\maketitle

\section{introduction}

Axion was originally proposed in connection with a solution to the strong CP problem in quantum chromodynamics (QCD) \cite{Peccei:1977ur,Peccei:2006as,Weinberg:1977ma, Wilczek:1977pj}. Another type of pseudo scalar particle similar to axions but with independent coupling and mass parameters is known as Axion-Like Particles (ALPs) \cite{Jaeckel:2010ni}. The existence of ALP has been predicted by theories beyond the Standard Model, such as string theory \cite{Svrcek:2006yi, Arvanitaki:2009fg, Marsh:2015xka}. Compared to axions, ALPs offer a broader parameter space for exploration.

The coupling between ALP and photon is extremely weak. However, astrophysical observations provide an excellent opportunity to detect ALP effects due to the long propagation distances involved. As photons travel through magnetic fields, they can oscillate into ALPs and and back along their trajectory \cite{Raffelt:1987im}. This ALP-photon oscillation effect can induce spectral irregularities that are potentially detectable by high energy $\gamma$-ray detectors. With recent advancements, experiments such as Fermi-LAT \cite{Fermi-LAT:2009ihh}, MAGIC \cite{MAGIC:2014zas}, HAWC \cite{HAWC:2020hrt}, and LHAASO \cite{LHAASO:2019qtb}, have achieved the capability to conduct precise measurements of the spectral energy distribution (SED) from high energy astrophysical sources, providing important data for ALP searches. Numerous studies have explored this phenomenon, yielding  a multitude of constraints on ALP parameters \cite{DeAngelis:2007dqd, Hooper:2007bq, Simet:2007sa, Mirizzi:2007hr, Belikov:2010ma, DeAngelis:2011id, Horns:2012kw, HESS:2013udx, Meyer:2013pny, Tavecchio:2014yoa, Meyer:2014epa, Meyer:2014gta, Fermi-LAT:2016nkz, Meyer:2016wrm, Berenji:2016jji, Galanti:2018upl, Galanti:2018myb, Zhang:2018wpc, Liang:2018mqm, Bi:2020ths, Guo:2020kiq, Li:2021gxs, Cheng:2020bhr, Liang:2020roo, Xia:2018xbt}.

Blazars, a subclass of active galactic nuclei (AGN), are widely studied in ALP searches due to their prominence in the high energy $\gamma$-ray extragalactic sky. These sources harbor supermassive black holes surrounded by accretion disks and generate relativistic jets oriented towards Earth. Photons emitted from the emission region of a blazar may undergo conversion into ALPs within the magnetic field of the jet and potentially revert back into photons within the Galactic magnetic field. However, many astrophysical magnetic fields are not precisely determined, leading to uncertainties in the constraints on ALP parameters. 

In the case of blazars, the primary source of uncertainty lies in the blazar jet magnetic field (BJMF). Many previous studies have employed a BJMF model dominated by a toroidal field \cite{Meyer:2013pny,Meyer:2014epa,Meyer:2014gta,Galanti:2018upl,Pant:2022ibi,Li:2020pcn,Li:2021gxs,Gao:2023dvn,Pant:2023omy}, where the field strength is proportional to the magnetic field strength in the emission region, denoted as $B_0$, and inversely proportional to the distance from the central black hole. Given that $B_0$ is determined through multi-wavelength spectral fitting and is correlated with other parameters, such as the jet’s Doppler factor, it is essential to simultaneously vary these parameters in the analysis.

The impact of variations in these parameters on ALP constraints is studied in Ref.~\cite{Gao:2024wpn}, where the multi-wavelength observations of Mrk 421, well-matched by the synchrotron self-Compton (SSC) model, are considered. 
Within the leptonic scenario \cite{Bu:2019qqg,Armando:2023zwz,Gao:2024wpn}, the high energy peak of the SED is attributed to inverse Compton scattering. However, the origin of this high energy peak remains under debate for some blazars, with hadronic models providing an alternative explanation. In the hadronic scenario, the high energy component can be attributed to proton synchrotron emission, or emissions from secondary particles generated in photohadronic (p$\gamma$) and hadronuclear (pp) interactions. Since leptonic and hadronic models predict different values for the magnetic field strength $B_0$, this introduces additional uncertainty in ALP constraints. Consequently, in this work, we aim to examine the differences in constraints derived from the leptonic and  hadronic models. 

In this study, we utilize the high energy $\gamma$-ray spectrum of Mrk 501 during its low-activity period, observed by Fermi-LAT and MAGIC from 2017-06-17 to 2019-07-23, to investigate the ALP-photon oscillation. Mrk 501, a blazar with a redshift of 0.034 \cite{MAGIC:2014rit}, is classified as a high synchrotron peaked BL Lac object. A study conducted by the MAGIC Collaboration in 2023 \cite{MAGIC:2022mhv} on Mrk 501 indicated that both leptonic and hadronic scenarios can adequately account for the data during this period, rendering  it a suitable target for our analysis.
Furthermore, we consider a different BJMF model as presented in Ref. \cite{Davies:2020uxn}. This model incorporates a tangled component and a helical component that turns from poloidal (aligned along the jet) to toroidal (transverse) as it propagates along the jet \cite{Prior:2019tnm,Davies:2020uxn,Murphy:2013zk}. The inclusion of these components would lead to different constraints compared to the conventional toroidal BJMF model.

This paper is organized as follows. In Sec. \ref{sec:ALPs and magnetic}, we introduce ALP photon oscillations and the astrophysical context of $\gamma$-ray photon propagation from Mrk 501. In Sec. \ref{sec:SSC}, we discuss the milit-wavelength SED fitting. The research methodologies utilized in this study is outlined in Sec. \ref{sec:method}. Our results are presented and elucidated in Sec. \ref{sec:result}.  Finally, we summarize our study in Sec. \ref{sec:summary}.

\section{ALP-photon oscillations and astrophysical environments}\label{sec:ALPs and magnetic}

In this section, we introduce the ALP-photon oscillation effect. We also introduce the astrophysical environment through which $\gamma$-ray photons originating from Mrk 501 traverse during their propagation.

\subsection{ALP-photon oscillations}\label{sec:ALPs}

The ALP Lagrangian describing the interaction between the ALP and photons can be written as
\begin{equation}
\mathcal{L}= \frac{1}{2}{\partial^\mu} a {\partial_\mu} a-\frac{1}{2}m_a^2 a^2+g_{a\gamma} a \boldsymbol{E} \cdot \boldsymbol{B},
\end{equation}
where $a$ denotes the ALP field, $m_a$ is the mass of the ALP, $g_{a\gamma}$ is the coupling parameter between the ALP and photon,
$\boldsymbol{E}$ denotes the electric field, and $\boldsymbol{B}$ denotes the magnetic field. In a reference frame where the photon propagation direction is along the $z$-axis and the transverse magnetic field $B_{t}$ aligns with the $y$-axis,   the propagation equation of the ALP-photon beam with energy $E \gg m_a$ can be expressed as
\begin{equation}\label{equ:transfer}
\left( i \frac{d}{dz} + E + \mathcal{M}_0 \right) \Psi(z) = 0,
\end{equation}
where $\Psi(z) = (A_1, A_2, a)^T$ represents the state vector of the beam, $A_1$ and $A_2$ denote the amplitudes of the photon field along the $x$-axis and $y$-axis, respectively. 
The mixing matrix $\mathcal{M}_0$ is given by
\begin{equation}\label{equ:mixmatrix}
    \mathcal{M}_0 = \begin{bmatrix}
\Delta_{\perp} & 0 & 0 \\
0 & \Delta_{\parallel} & \Delta_{a\gamma} \\
0 & \Delta_{a\gamma}  & \Delta_{a} 
\end{bmatrix},
\end{equation}
with {$\Delta_{\perp} = \Delta_{pl} +2\Delta_{\rm QED} + \chi_{\rm CMB}$, $\Delta_{\parallel} = \Delta_{pl} + 7/2\Delta_{\rm QED} + \chi_{\rm CMB}$}, $\Delta_{a}=-m_a^2/(2E)$, and $\Delta_{a\gamma}=g_{a\gamma}B_t/2$. Here, $\Delta_{pl} = -\omega_{\rm pl}^2/(2E)$ represents the plasma contribution, and the plasma frequency is $\omega_{\rm pl} = ( {4\pi e^2n_e}/{m_e})^{{1}/{2}}$. The QED vacuum polarization is given by $\Delta_{\rm QED}=\alpha E/(45\pi)(B_{\perp}/B_{\rm cr})^2$, where $B_{\rm cr}$ is the critical magnetic field. {$\chi_{\rm CMB}=44\alpha^2\rho_{\rm CMB} /(135 m_{e}^4)$ quantifies photon-photon dispersion induced by the cosmic microwave background (CMB), with $\rho_{\rm CMB}$ the CMB energy density \cite{Dobrynina:2014qba}.}

The probability that a photon polarized along the $y$ axis converts into an ALP after a propagation $L$ in a constant magnetic field can be explicitly given by
\begin{equation}\label{equ:probga}
P_{\gamma \rightarrow a} = \sin^2 (2 \theta) \sin^2 \left( \frac{\Delta_{\textrm{osc}}L}{2} \right),
\end{equation}
where $\theta$ represents the ALP-photon mixing angle
\begin{equation}\label{equ:mixangle}
\theta = \frac{1}{2} \arctan \left(  \frac{2\Delta_{a\gamma}}{\Delta_{\parallel} - \Delta_a } \right),
\end{equation}
and $\Delta_{\textrm{osc}}$ represents the oscillation wave number
\begin{equation}\label{equ:oscwn}
\Delta_{\textrm{osc}} = [(\Delta_{\parallel} - \Delta_a )^2 + 4 \Delta_{a\gamma}^2 ]^{1/2}.
\end{equation}

To obtain the exact state of the ALP-photon beam, we should numerically solve the propagation equation Eq.~\ref{equ:transfer}.
The is equation can be rewritten as a Von Neumann-like equation \cite{DeAngelis:2007dqd,Mirizzi:2009aj}:
\begin{equation}\label{equ:von Neumannn-like}
    i\frac{d\rho}{dz} = [\rho, \mathcal{M}_0],
\end{equation}
where $\rho \equiv \Psi \otimes \Psi^\dagger$ is the density matrix of the beam.
The solution to Eq. \ref{equ:von Neumannn-like} can be expressed in terms of a transfer matrix $\mathcal U(z,z_{0})$, determined by the eigenvalues of $\mathcal{M}_0$, as detailed in Ref.~\cite{Galanti:2022ijh}.
Therefore, the beam state at position $z$ can be written as $\rho = \mathcal U(z,z_{0}) \rho(z_0) \mathcal U^\dagger(z,z_{0})$. The probability for a photon to transition from state $\rho_{0}$ to state $\rho_{1}$ is given by \cite{Galanti:2022ijh} 
\begin{equation}\label{equ:Pgaga}
P_{\gamma\gamma} = \mathrm{Tr}\left(\rho_{1} \mathcal U(z,z_{0}) \rho_{0} \mathcal U^{\dagger}(z,z_{0})\right).
\end{equation}

Considering the challenges in measuring the polarization of high energy $\gamma$ rays, it is common to assume that the high energy $\gamma$ rays emitted from the source are unpolarized. Therefore, the initial state $\rho_0$ is taken to be 
$\text{diag}(1/2,1/2,0)$.

\subsection{Astrophysical environments}\label{sec:magnetic}

High energy photons from the blazar are emitted from a region close to the black hole at a distance $r_{\rm VHE}$. The magnetic field strength in this region is denoted by $B_0$. The relationship between the photon energy in the jet’s frame $E_j$, and the photon energy in the laboratory frame $E_L$, is given by $E_j = E_L/\delta_D$, where $\delta_D$ is the Doppler factor. 
As these photons traverse through the jet, the phenomenon of ALP-photon oscillations can occur within the BJMF. We consider two different models for the BJMF to investigate the  oscillation effects and potential constraints on ALP parameters.

In the context of the first BJMF model (referred to as the toroidal model), which has been widely utilized in previous studies to investigate the ALP effects, the influence of the poloidal magnetic field component diminishes as the distance from the central black hole increases. This allows the toroidal magnetic field to emerge as the dominant component in the jet environment \cite{Pudritz:2012xj, Begelman:1984mw}. The toroidal magnetic field distribution along with an electron density distribution \cite{Pudritz:2012xj, Begelman:1984mw, OSullivan:2009dsx} can be described as 
\begin{equation}
    B^{{\rm jet}}(r) = B_0^{\rm jet}  (r/r_{\rm VHE})^{-1},
\label{BJMF}
\end{equation}
\begin{equation}
    n^{\rm jet}_{\rm el}(r) = n_0^{\rm jet} (r/r_{\rm VHE})^{-2},
\end{equation}
where $r_{\rm VHE}$ represents the distance between the emission region and the central black hole, and $n_0^{\rm jet}$ denote the electron density in the emission region. The values of these parameters can be determined through the subsequent multi-wavelength SED fitting analyses.

The second model (referred to as the helical+tangled model)  incorporates both a helical component and a tangled component \cite{Davies:2020uxn}. The helical magnetic field transitions from being aligned with the jet direction to a transverse orientation as it propagates. The strength of this transverse magnetic field diminishes with distance along the jet, and is characterized by the relationship  $B_T \propto r^{-\alpha}$. Due to uncertainties regarding the precise rate of this transition, we simplify the analysis by assuming that the helical magnetic field undergoes this transition at a specific distance denoted as $r_T$. The relationship between the strengths of the two components can be given by \cite{Murphy:2013zk}:
\begin{equation}
    \frac{B_{\text{tangled}}^2}{B_{\text{helical}}^2} = \frac{f}{1-f},
\label{jetHT}
\end{equation}
where $f$ represents the proportion of the tangled magnetic field in the total magnetic energy density.  

After high energy photons leave the jet region and the host galaxy, they propagate into extragalactic space where the magnetic field strength is relatively weak and not precisely determined (typically $\lesssim 10^{-9}$ G) \cite{Kronberg:1993vk,Blasi:1999hu,Durrer:2013pga}. In this extragalactic region, {we do not consider ALP-photon oscillations induced by the extragalactic magnetic field and focus on the absorption of high-energy photons by the extragalactic background light (EBL) via the pair-production process $\gamma + \gamma_{\rm EBL} \to e^+ + e^-$.} When examining the EBL absorption effects, we employ the model from Ref. \cite{Franceschini:2008tp,MAGIC:2022mhv}. {Regarding the Milky Way's magnetic field, we use the fiducial "base" model provided in Ref. \cite{Unger:2023lob}, this model comprises three components: a spiral disk field with three-mode grand-design structure, an explicit toroidal halo field, and a coasting X-field featuring logistic-sigmoid radial dependence. This configuration assumes no correlation between thermal electron density and magnetic field strength, and incorporates striation as a free parameter. Rotation measure (RM) and linear polarization (Q/U) sky maps are generated using the YMW16 electron density model \cite{Yao:2017kcp} and cosmic-ray electrons distributed at 6 kpc scale height. Additionally, we derive constraints on ALP using the magnetic field model from Ref. \cite{Jansson:2012rt}, finding results consistent with those obtained from the "base" model in Ref. \cite{Unger:2023lob}.} 

Considering the combined effects of the various processes discussed above, the observed energy spectrum of the target source $\Phi_{\rm obs}$  can be expressed as
\begin{equation}
\Phi_{\rm obs} = P_{\gamma\gamma}\Phi_{\rm in},
\end{equation}
where  $\Phi_{\rm in}$ is the intrinsic energy spectrum of the source, and $P_{\gamma\gamma}$ represents the survival probability of high energy photons reaching the Earth. In this study, we use the package gammaALPs \cite{Meyer:2021pbp} to calculate $P_{\gamma\gamma}$.

\section{Multi-wavelength  fitting and magnetic field strength in the emission region}\label{sec:SSC}

\begin{figure} \centering \includegraphics[width=0.45\textwidth]{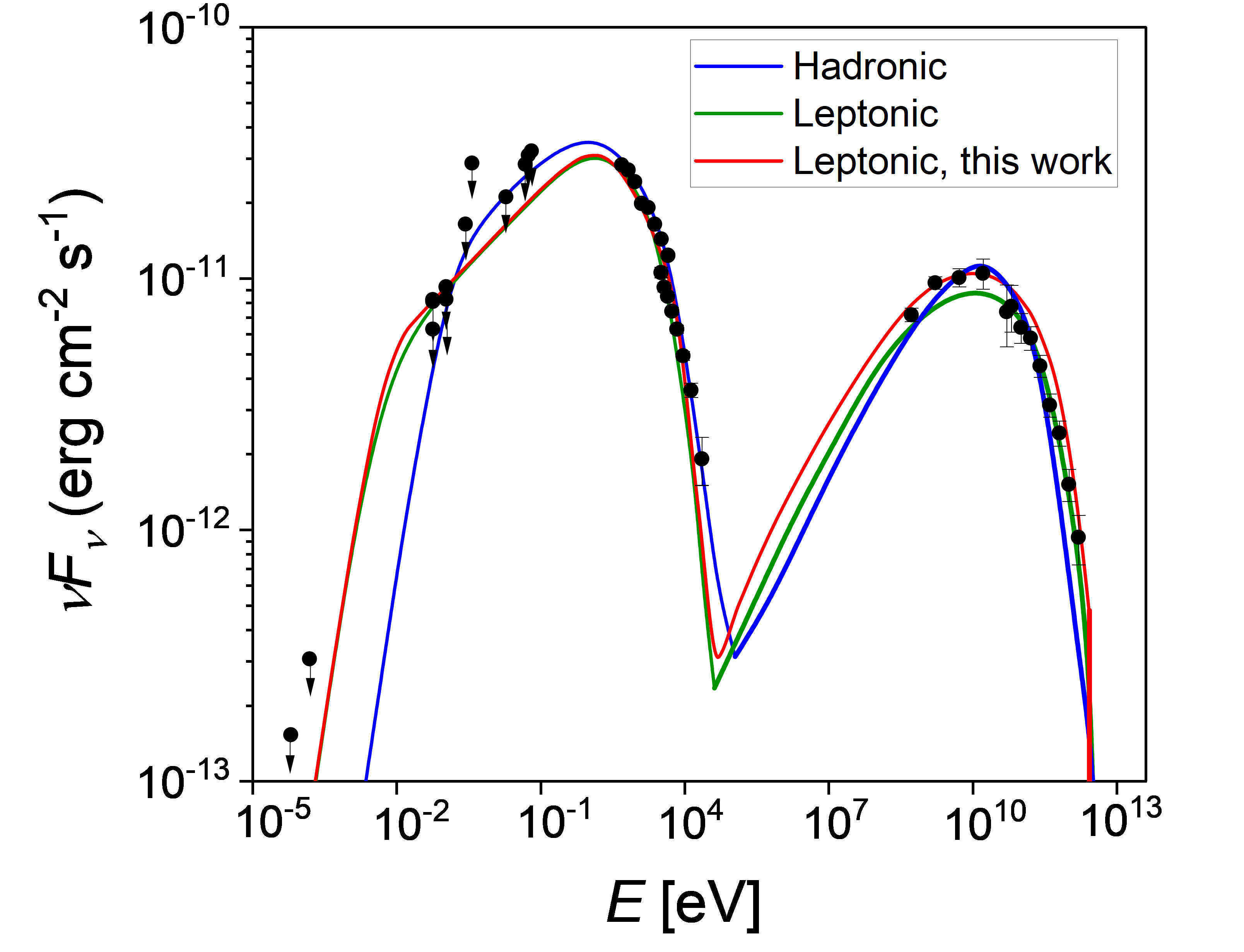} 
\caption{\raggedright The multi-wavelength SED and fitting results. The red solid line represents the best-fit result for the leptonic scenario in our work. For comparison, the results for the hadronic and leptonic scenarios in Ref. \cite{MAGIC:2022mhv} are also shown.} 
\label{fig:multi_SEDs_fit} 
\end{figure}

The multi-wavelength SED of blazars commonly exhibits a double-peak structure. The lower-energy peak is attributed to synchrotron radiation emitted by relativistic electrons, while the origin of the higher-energy peak remains a subject of ongoing research and debate. In the scenario where leptonic processes dominate, the higher peak is ascribed to SSC emission. This mechanism involves the upscattering of synchrotron photons by relativistic electrons through inverse Compton scattering within the jet, leading to the production of high energy photons. 

On the other hand, in the hadronic scenario, the higher peak is associated with processes involving relativistic protons. This can occur through synchrotron radiation emitted by the protons themselves or through secondary particles generated by interactions between high energy protons and photons within the jet.
While hadronic models for luminous blazars often require super-Eddington luminosities, which pose significant energy constraints, this may not be the case for high-synchrotron-peaked BL Lac objects, especially during low-flux states. The hadronic emission mechanism may not account for the high variability observed in Mrk 501 \cite{Galanti:2023uam}. However, in the low-activity phase under consideration, the low flux variability still enables the hadronic emission mechanism to explain the observation of Mrk 501.

Both leptonic and hadronic models are capable of effectively describing the multi-wavelength SED of Mrk 501 during its low-activity phase, from 2017-06-17 to 2019-07-23, as observed by several experiments, including MAGIC, Fermi-LAT, NuSTAR, Swift, GASP-WEBT, and OVRO \cite{MAGIC:2022mhv}. Here, we briefly introduce these two scenarios and present the fitting results derived from their application. 

In the one-zone SSC model, photons are emitted from a compact, spherical region situated in the vicinity of the central black hole. This emission region, characterized by  a radius $R$, contains a population of relativistic electrons immersed in a uniform magnetic field with strength $B_0$. The energy distribution of these relativistic electrons is described by a broken power law
\begin{equation}
N(\gamma) = 
\begin{cases} 
N_{0} \cdot \gamma^{-\alpha_1}, & \text{if } \gamma_{\rm min}< \gamma < \gamma_{\rm break} \\
N_{0} \cdot \gamma^{-\alpha_2} \gamma_{\rm break}^{\alpha_2-\alpha_1}, & \text{if }
\gamma_{\rm break}< \gamma < \gamma_{\rm max},
\end{cases}
\end{equation}
where $N_{0}$ is the normalization constant, and $\alpha_1$ and $\alpha_2$ represent the spectral indices below and above the break energy, respectively. The parameters $\gamma_{\rm min}$, $\gamma_{\rm break}$, and $\gamma_{\rm max}$ denote the minimum, break, and maximum Lorentz factors of the electron population, respectively.

The emission region moves along the jet with a bulk Lorentz factor $\Gamma$. This motion establishes a connection between the photon energy in the frame of the emission region and the energy observed in the laboratory frame, which is characterized by the Doppler factor  $\delta_{D}$. This factor is given by $\delta_{D} = [\Gamma (1 - \beta \cos \theta)]^{-1}$, where $\beta$ represents the velocity of the emission region. Assuming an angle $\theta = 1/\Gamma$ between the jet axis and the line of sight, the expression for the Doppler factor simplifies to  $\delta_{D} = \Gamma$.

In Ref. \cite{MAGIC:2022mhv}, 
the process of fitting the SED of Mrk 501 is conducted in two stages. Initially, the fitting is performed using naima \cite{Zabalza:2015bsa} for an initial assessment. This is followed by a more refined fitting procedure using JetSet \cite{Tramacere:2011qw}, where \( B_0 \) is held fixed. In our investigation, we utilize the packages agnpy  \cite{Nigro:2021pxy} and sherpa \cite{Freeman:2001uc} for the SED fitting process, and allow the magnetic field strength to vary freely during the fitting procedure. The values of the emission region radius \( R \), minimum Lorentz factor \( \gamma_{\rm min} \), and Doppler factor \( \delta_{D} \) are adopted to be  \( 1.14 \times 10^{17} \) cm, 1000, and 11, respectively. Based on the assumption \( r_{\rm VHE} = R/\theta \), it can be obtained
that $r_{\rm VHE}$ = \( 1.25 \times 10^{18} \) cm. The resulting SED, represented by the red solid line in Fig. \ref{fig:multi_SEDs_fit} and characterized by the parameters detailed in Tab. \ref{tab: multi_SEDs_params}, indicates that the leptonic scenario offers a reliable explanation of the observational data. Our fitting results are in agreement with those reported in Ref. \cite{MAGIC:2022mhv}, as demonstrated by the green solid line in Fig. \ref{fig:multi_SEDs_fit}.

In the hadronic scenario, the high energy emissions primarily originate from relativistic protons rather than electrons. These emissions can be attributed to two main mechanisms: synchrotron radiation emitted by protons in strong magnetic fields, and interactions between protons and surrounding photons, leading to the production of secondary particles, such as pions, which subsequently decay into photons.
In this scenario, the emission region consists of both relativistic electrons and protons, each following simple power-law distributions. The emission radius and Doppler factor remain fixed,  similar to the one-zone SSC model, while the minimum Lorentz factors are set to be 400 for electrons and 1 for protons. The fitting results in the hadronic scenario, as detailed in Ref. \cite{MAGIC:2022mhv}, are depicted by the blue solid line in Fig. \ref{fig:multi_SEDs_fit}. These results underscore a satisfactory agreement with the observational data, with the corresponding parameter values outlined in Tab. \ref{tab: multi_SEDs_params}.

\begin{table}[htbp]
  \centering
  \caption{\raggedright The  parameters for the leptonic and hadronic emission scenarios adopted in our analysis. The parameter values for the leptonic scenario are derived from the multi-wavelength fitting using the package agnpy. The parameter values for the hadronic scenario are taken from Ref. \cite{MAGIC:2022mhv}.}
  \label{tab: multi_SEDs_params}
  \renewcommand{\arraystretch}{1.5}
  \setlength{\tabcolsep}{15pt}
  \begin{tabular}{ccc}
    \hline
    \hline
    parameter & Leptonic & Hadronic   \\ 
    \hline
    B(G) & 0.029 &  3 \\
    R(cm) & $1.14 \times 10^{17}$ & $1.14 \times 10^{17}$  \\
    $\delta_{D}$ & 11 & 11  \\
    $\alpha_{\rm e,1}$ & 2.7 & 2.5  \\
    $\alpha_{\rm e,2}$ & 3.6 & -  \\
    $\alpha_{\rm p}$ & - & 2.2  \\
    $\gamma_{\rm min,e}$ & 1000 & 400  \\
    $\gamma_{\rm break,e}$ & $2.4 \times 10^{5}$ & - \\
    $\gamma_{\rm max,e}$ & $1.2 \times 10^{6}$ & $3.5 \times 10^{4}$  \\
    $\gamma_{\rm min,p}$ & - &  1 \\
    $\gamma_{\rm max,p}$ & - &  $1.1 \times 10^{10}$ \\
    \hline
  \end{tabular}
\end{table}

The fitting spectra indicate that both the hadronic and leptonic scenarios provide viable explanations for the SED of the source. However, a notable distinction arises in the estimated best-fit magnetic field strength \( B_0 \), with the hadronic scenario suggesting a value approximately two orders of magnitude higher than that inferred from the leptonic model. This discrepancy highlights the distinct characteristics of the two scenarios in describing the emission processes of the blazar. Currently, observational data do not definitively favor one model over the other, underscoring the need for future neutrino observations, which  have the potential to discriminate the hadronic scenario from alternative scenarios. 

\section{method}\label{sec:method}

This section outlines the methods used for fitting the energy spectrum and the statistical procedures employed to constrain the ALP parameters. We utilize high energy $\gamma$-ray spectra observed from Mrk 501 by Fermi-LAT and MAGIC in the analysis.
We consider four intrinsic spectral forms for fitting the SED: a power law with sub-/super-exponential cutoff (SEPWL), a log-parabola (LP), a log-parabola with an exponential cutoff (ELP), and a power law with an exponential cutoff (EPWL). The selection of the spectral form is based on its fitting performance, with the spectral free parameters denoted as $\Theta$. A $\chi^2$ fit is conducted using the spectral data obtained from Fermi-LAT and MAGIC, where the $\chi^2$ structure is defined as
\begin{align}
    \chi^2(\Theta; m_a, g_{a\gamma}) &= \sum\limits_{i=1}^{M} \frac{(\Phi(E_i) |_{\rm LAT} - \Phi_{{\rm obs},i} |_{\rm LAT})^2}{\delta_{{\rm obs},i} |_{\rm LAT}^2} \notag \\
    & + \sum\limits_{i=1}^{N} \frac{(\Phi(E_i) |_{\rm MAGIC} - \Phi_{{\rm obs},i} |_{\rm MAGIC})^2}{\delta_{{\rm obs},i} |_{\rm MAGIC}^2},
\end{align}
where $i$ represents the $i$-th energy bin, $\Phi(E_i)$ is the expected average flux in the $i$-th energy bin, $\Phi_{{\rm obs},i}$ is the observed flux value, and $\delta_{{\rm obs},i}$ is the flux uncertainty in the observation, and $M$ and $N$ represent the number of data bins for Fermi-LAT and MAGIC, respectively.

To constrain the ALP parameters, we define the test statistic (TS) as the difference in $\chi^2$ values between the ALP hypothesis and the null hypothesis:
\begin{equation}\label{equ:TS_CLs}
    {\rm TS}(m_a, g_{a\gamma}) = \chi_{\rm ALP}^2(\hat{\hat{\Theta}}; m_a, g_{a\gamma}) - \chi^2_{\rm Null}(\hat{\Theta}),
\end{equation}
where \(\chi^2_{\rm ALP}(\hat{\hat{\Theta}}; m_a, g_{a\gamma})\) is the best-fit \(\chi^2\) value under the ALP hypothesis with parameters \((m_a, g_{a\gamma})\),  \(\chi^2_{\rm Null}\) is the best-fit \(\chi^2\) value under the null hypothesis, and \(\hat{\hat{\Theta}}\) and \(\hat{\Theta}\) represent the best-fit parameters of the intrinsic spectrum under the ALP and null hypotheses, respectively.
The non-linear relationship between the ALP parameters and the modifications to the photon spectrum would result in the TS not following the standard $\chi^2$ distribution. Consequently, the conventional Wilks’ theorem \cite{wilk} cannot be directly applied in this analysis  \cite{Fermi-LAT:2016nkz}. To set the constraints on ALP parameters by evaluating the TS distribution, we employ the $\rm CL_s$ method \cite{Junk:1999kv, Read:2002hq_cls, Lista:2016chp}. This method has been used in utilized in studies focusing on detecting ALP effects in high energy $\gamma-$ray spectra in   \cite{Gao:2023dvn, Li:2024ivs, Gao:2023und, Gao:2024wpn}, with a detailed description referenced in Ref. \cite{Gao:2023dvn}.

\section{results}\label{sec:result}

\begin{figure} \centering \includegraphics[width=0.45\textwidth]{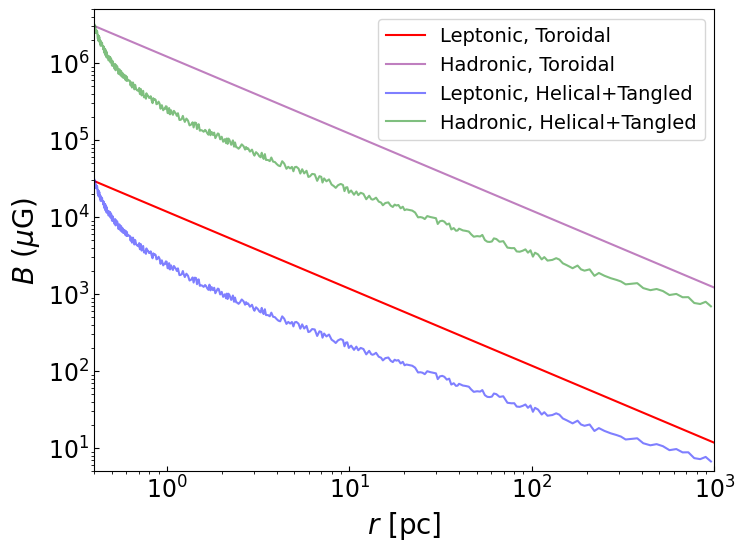} 
\caption{\raggedright
The transverse magnetic field distributions in the blazar jet. The red and blue solid lines represent the two BJMF configurations in the leptonic scenario, while the purple and green solid lines correspond to the hadronic scenario. The red (purple) and blue (green) lines represent the distributions in toroidal and helical+tangled models, respectively.} 
\label{fig:B_r} 
\end{figure}

In this section, we present the constraints on the ALP parameters derived from the observations of Mrk 501 by MAGIC and Fermi-LAT from 2017-06-17 to 2019-07-23. The SED of Mrk 501 can be effectively explained by both leptonic and hadronic emission scenarios, thus prompting us to consider both scenarios in the analysis. Furthermore, we utilize two BJMF models to investigate ALP-photon oscillations.
In the toroidal BJMF model, the parameters are set as discussed in Sec. \ref{sec:SSC}. In the helical+tangled BJMF model, three additional parameters are introduced compared to the former model, denoted as $\alpha$, $r_T$, and $f$. Previous simulation studies on Mrk 501 suggest that the parameter $\alpha$ ranges from 0.2 to 1.5, $r_T$ ranges from 0.1 to 10 pc, and $f$ ranges from 0 to 0.7 \cite{Davies:2020uxn}. Given the uncertainty in precisely determining these parameters, we adopt specific values of $\alpha = 1$, $r_T = 0.3$ pc, and $f = 0.3$, in alignment with the values used in Ref. \cite{Meyer:2021pbp}. In this study, we consider four jet magnetic field configurations: the value of $B_0$ derived from leptonic or hadronic scenarios, and either the toroidal or helical+tangled BJMF model. The distributions of transverse magnetic field within the jet corresponding to these four cases are displayed in Fig. \ref{fig:B_r}. 

\begin{figure*}[ht]
  \centering
  \subfloat[Leptonic, Toroidal]{\includegraphics[width=0.45\textwidth]{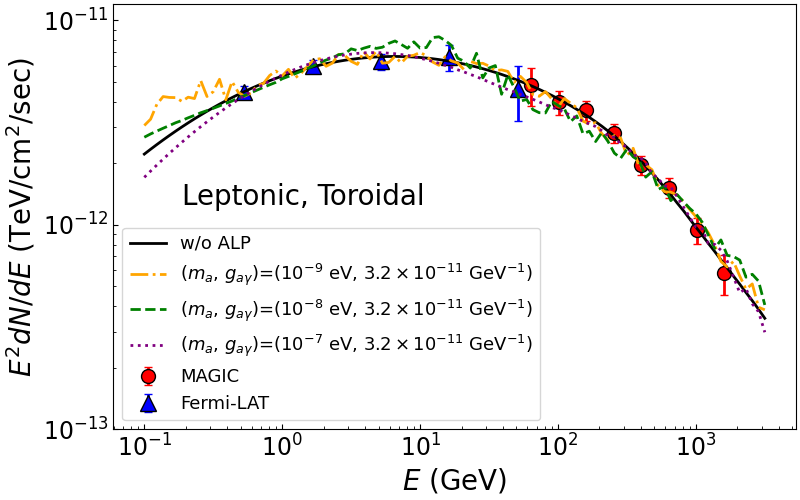}\label{fig:fit_leptonic}}
  \hspace{0.05\textwidth} 
  \subfloat[Hadronic, Toroidal]{\includegraphics[width=0.45\textwidth]{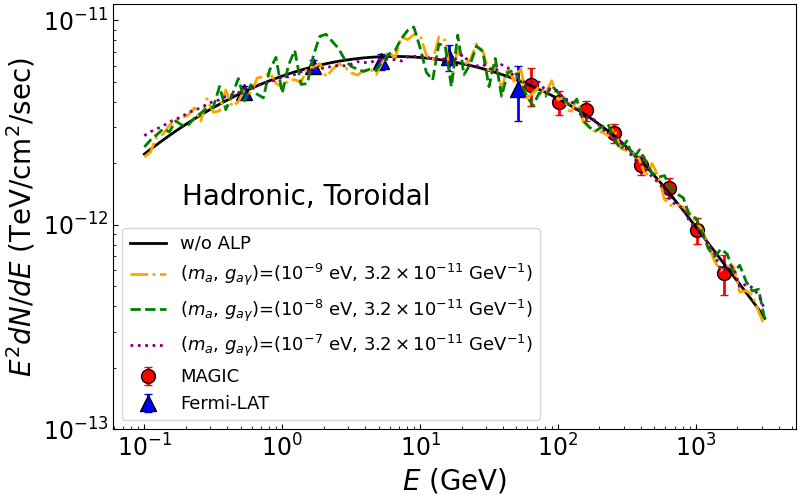}\label{fig:fit_hadronic}}

  \subfloat[Leptonic, Helical+Tangled]{\includegraphics[width=0.45\textwidth]{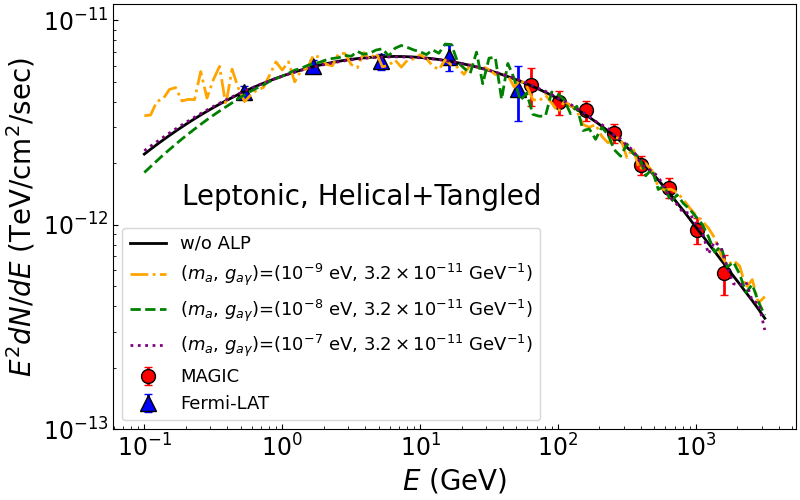}\label{fig:fit_leptonic_JHT}}
  \hspace{0.05\textwidth}  
  \subfloat[Hadronic, Helical+Tangled]{\includegraphics[width=0.45\textwidth]{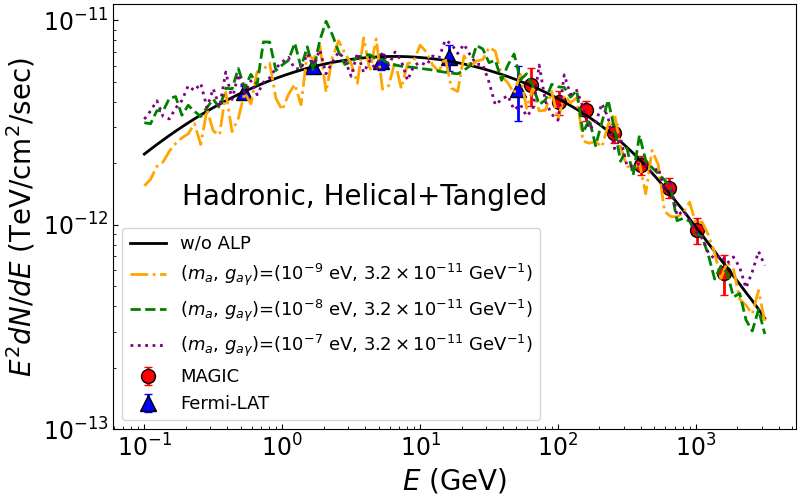}\label{fig:fit_hadronic_JHT}}

  \caption{\raggedright The  high energy $\gamma-$ray spectra under the null and ALP hypotheses. Four subfigures correspond to four magnetic field cases. The red and blue points represent the spectrum data of Mrk 501 observed by MAGIC and Fermi-LAT from 2017-06-17 to 2019-07-23. The black solid line  represents the best-fit spectrum under the null hypothesis. The orange, green, and purple dashed lines represent best-fit spectra for three ALP parameter points.}
  \label{fig:fit}
\end{figure*}

\begin{figure*}[ht]
  \centering
  \subfloat[Leptonic, Toroidal]{\includegraphics[width=0.4\textwidth]{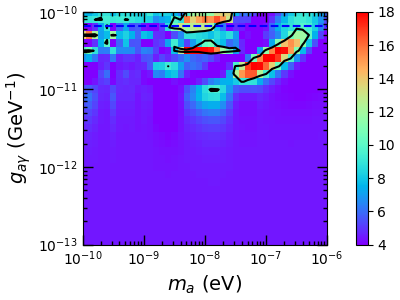}\label{fig:leptonic}}
  \hspace{0.05\textwidth}  
  \subfloat[Hadronic, Toroidal]{\includegraphics[width=0.4\textwidth]{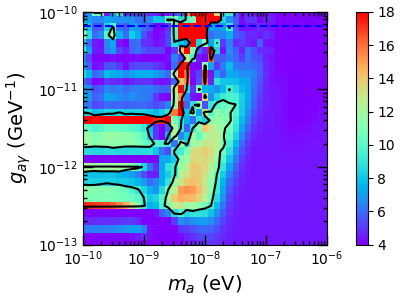}\label{fig:hadronic}}
  
  \subfloat[Leptonic, Helical+Tangled]{\includegraphics[width=0.4\textwidth]{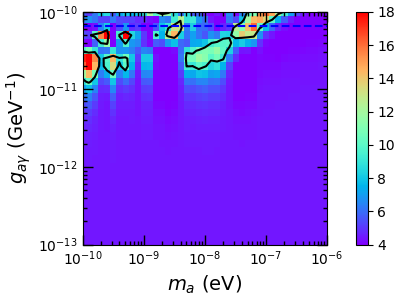}\label{fig:leptonic_JHT}}
  \hspace{0.05\textwidth}  
  \subfloat[Hadronic, Helical+Tangled]{\includegraphics[width=0.4\textwidth]{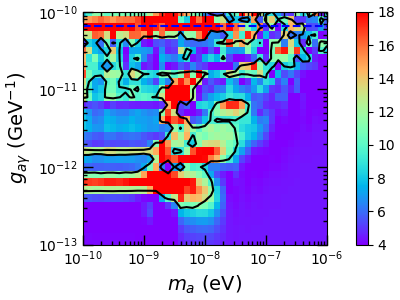}\label{fig:hadronic_JHT}}
  \caption{\raggedright The distribution of $\chi^2$ in the $m_a-g_{a\gamma}$ plane for four magnetic field configurations. The black solid lines represent the 95\% C.L. constraints derived in this study. The blue dashed line represents the CAST constraint \cite{CAST:2017uph}.}
  \label{fig:CLs}
\end{figure*}

The selection of the intrinsic spectrum form is determined by the Akaike Information Criterion (AIC) \cite{Akaike:1974vps}, which is defined as \( \text{AIC} = 2k - 2 \ln \hat{L} \), where \( \hat{L} = \exp\left(-\chi^2(\hat{\Theta})/2\right) \) is the maximized likelihood function, and $k$ denotes the number of free parameters. For the SEPWL, LP, ELP, and EPWL spectral forms, the numbers of free parameters are 4, 3, 4, and 3, respectively. The AIC values for the SEPWL, LP, ELP, and EPWL spectral forms are { 11.1, 10.4, 11.0, and 32.7, respectively. Among these, the LP model yields the smallest AIC, and is thus adopted for the analysis.}

The fitting spectra under the null hypothesis align well with the data, as shown in Fig. \ref{fig:fit}. We also present the fitting results for three benchmark parameter points under the ALP hypothesis, with the corresponding results depicted by the orange, green, and purple dashed lines in Fig. \ref{fig:fit}. Each of the four subfigures corresponds to different magnetic field configurations. 
We perform a scan across the parameter space for the ALP mass \( m_a \) ranging from \( 10^{-10} \) eV to \( 10^{-6} \) eV, and the ALP-photon coupling constant \( g_{a\gamma} \) ranging from \( 10^{-13} \) to \( 10^{-10} ~\rm {GeV}^{-1} \), utilizing logarithmic binning at an interval of 0.1. Each parameter point of \((m_a, g_{a\gamma})\) undergoes evaluation for exclusion based on the $\rm CL_s$ method. The 95\% confidence level (C.L.) constraints on the parameter region are established, with the resulting constraints depicted as black solid lines in Fig. \ref{fig:CLs}, superimposed on a color-coded heat map illustrating the \(\chi^2\) value for each parameter point of \((m_a, g_{a\gamma})\) under the ALP hypothesis.

In comparison to the results in the leptonic scenario, as  illustrated in Fig.~\ref{fig:leptonic} and \ref{fig:leptonic_JHT}, the hadronic scenario imposes significantly more stringent constraints on the ALP parameters, as evidenced in Fig.~\ref{fig:hadronic} and Fig.~\ref{fig:hadronic_JHT}. These discrepancies are primarily attributed to the substantial difference in the magnetic field strength $B_0$ between the two scenarios. The value of $B_0$ in the hadronic scenario is two orders of magnitude greater than that in the leptonic scenario, as visually depicted by the solid lines in Fig. \ref{fig:B_r}. In the hadronic scenario and the toroidal model, the constraints are most stringent, reaching approximately \( g_{a\gamma} \sim 2.5 \times 10^{-13} ~\rm {GeV}^{-1} \) for an ALP mass \( m_a \) around {\( 4.0 \times 10^{-9} \, \mathrm{eV} \)}.

The two BJMF models exhibit different constraint regions, where the toroidal model imposes more rigorous limits on \( g_{a\gamma} \) within the mass range of \( m_a \sim 10^{-7} - 10^{-6} \) eV. The differences in constraints between the two BJMF models can be understood by comparing the transverse magnetic field distributions within the jet, particularly for \( r > 1 \, \mathrm{pc} \), as depicted in Fig. \ref{fig:B_r}. The magnetic field strengths in the two BJMF models for the same emission scenario exhibit a discrepancy of nearly an order of magnitude, leading to variations in the resulting constraints.

The complex behavior of the constraints in the hadronic scenario can be understood by Eqs.~\ref{equ:probga}, \ref{equ:mixangle}, and \ref{equ:oscwn}. In the hadronic scenario involving a strong magnetic field strength, the QED term $\Delta_{\rm QED}$ may exceed the ALP mass term $\Delta_a$ in the mixing matrix for small ALP masses below $\sim 10^{-9}$ eV. In such cases, the ALP mass in  $\Delta_a$ would not significantly impact the ALP-photon oscillation rate, as indicated by Eqs.~\ref{equ:probga}, \ref{equ:mixangle}, and \ref{equ:oscwn}. This suggests that the constraints in this mass region are not strongly dependent on the ALP mass. Furthermore, in this mass region, the mixing term $\Delta_{a\gamma} \sim g_{a\gamma} B$, which may be dominant in the mixing matrix Eq. \ref{equ:mixmatrix}, could play a crucial role in the oscillation probability. Consequently, the constraints on $g_{a\gamma}$ would significantly change due to the  oscillatory behavior outlined in Eq.~\ref{equ:probga}. When the ALP mass approaches $\sim 10^{-8}$ eV, the situation becomes more complex. This complexity arises as the contributions from $\Delta_{\rm QED}$ and $\Delta_a$ may become comparable, and these terms could become non-negligible compared to $\Delta_{a\gamma}$ for small $g_{a\gamma}$. In such intricate cases, the constraints exhibit a highly complex and nuanced behavior.

\begin{figure} \centering \includegraphics[width=0.45\textwidth]{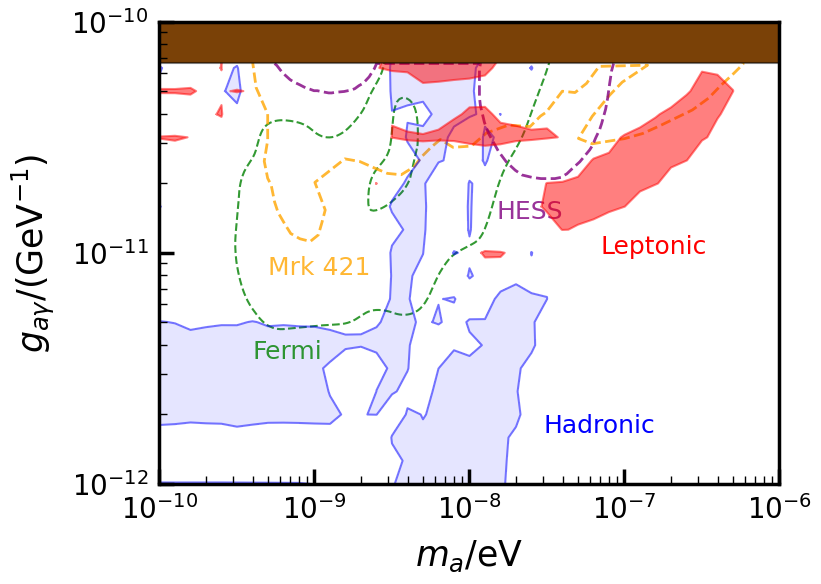}
\caption{\raggedright  The 95\% C.L. constraints derived under the leptonic (red contour) and hadronic (blue contour) scenarios for the toroidal BJMF model in this study. Comparative results include constraints from CAST (ochre contour) \cite{CAST:2017uph}, NGC 1275 observations by Fermi-LAT (green line) \cite{Fermi-LAT:2016nkz}, PG 1553+113 observations by H.E.S.S. II and Fermi-LAT (purple line) \cite{HESS:2013udx}, and Mrk 421 observations by ARGO-YBJ and Fermi-LAT (orange line) \cite{Li:2020pcn} (see also \cite{AxionLimits}).} 
\label{fig:com} 
\end{figure}

In Fig. \ref{fig:com}, we present the 95\% C.L. constraints derived from the toroidal model under both the leptonic and hadronic scenarios. For comparison, we include notable constraints from previous studies, such as the constraints from CAST \cite{CAST:2017uph}, observations of NGC 1275 by Fermi-LAT \cite{Fermi-LAT:2016nkz}, observations of PG 1553+113 by H.E.S.S. II and Fermi-LAT \cite{HESS:2013udx}, and observations of Mrk 421 by ARGO-YBJ and Fermi-LAT \cite{Li:2020pcn}. Our results indicate that, for both the leptonic and hadronic scenarios, the derived constraints complement those from earlier investigations in specific regions of the parameter space, with a notable enhancement in the case of the hadronic scenario. 

\section{conclusion}\label{sec:summary}

In this study, we investigate the ALP-photon oscillation effect using the SED of Mrk 501, as observed by MAGIC and Fermi-LAT during its low-activity period spanning from 2017-06-17 to 2019-07-23. We establish the constraints on the ALP parameter space based on both leptonic and hadronic emission scenarios for Mrk 501, and two BJMF models: a toroidal model and a model incorporating both helical and tangled components.

Our findings unveil notable disparities in the constraints derived from the leptonic and hadronic scenarios, as well as between the two BJMF models. Specifically, the hadronic scenario, characterized by a more strong magnetic field strength, imposes stricter limitations on the ALP parameters. Furthermore, at distances exceeding 1 pc, the magnetic field strength in the two BJMF models diverges by almost an order of magnitude, thereby accounting for the observed variability in constraints. We compare our outcomes with those of prior studies, revealing that our results complement earlier works, particularly within the hadronic scenario.

Looking forward, experiments such as LHAASO \cite{LHAASO:2019qtb} and CTA \cite{CTAConsortium:2013ofs}, which promise more precise measurements of very high energy \(\gamma\)-ray spectra, are expected to provide even more stringent constraints on the ALP parameter space. Furthermore, while both the leptonic and hadronic scenarios could explain the multi-wavelength observations, it is crucial to to ascertain the veracity of one over the other. Neutrino observations could offer valuable insights  in distinguishing between these scenarios, thereby presenting a pivotal test for ALP studies.

\acknowledgments
 The spectral data used in this paper are from Ref. \cite{MAGIC:2022mhv}, which are available at the website \url{https://cdsarc.cds.unistra.fr/viz-bin/cat/J/ApJS/266/37#/browse}. This work is supported by the National Natural Science Foundation of China under grant Nos. 12475057 and 12175248.

\bibliographystyle{apsrev}

\begin{thebibliography}{79}
\expandafter\ifx\csname natexlab\endcsname\relax\def\natexlab#1{#1}\fi
\expandafter\ifx\csname bibnamefont\endcsname\relax
  \def\bibnamefont#1{#1}\fi
\expandafter\ifx\csname bibfnamefont\endcsname\relax
  \def\bibfnamefont#1{#1}\fi
\expandafter\ifx\csname citenamefont\endcsname\relax
  \def\citenamefont#1{#1}\fi
\expandafter\ifx\csname url\endcsname\relax
  \def\url#1{\texttt{#1}}\fi
\expandafter\ifx\csname urlprefix\endcsname\relax\def\urlprefix{URL }\fi
\providecommand{\bibinfo}[2]{#2}
\providecommand{\eprint}[2][]{\url{#2}}

\bibitem[{\citenamefont{Peccei and Quinn}(1977)}]{Peccei:1977ur}
\bibinfo{author}{\bibfnamefont{R.~D.} \bibnamefont{Peccei}} \bibnamefont{and}
  \bibinfo{author}{\bibfnamefont{H.~R.} \bibnamefont{Quinn}},
  \bibinfo{journal}{Phys. Rev. D} \textbf{\bibinfo{volume}{16}},
  \bibinfo{pages}{1791} (\bibinfo{year}{1977}).

\bibitem[{\citenamefont{Peccei}(2008)}]{Peccei:2006as}
\bibinfo{author}{\bibfnamefont{R.~D.} \bibnamefont{Peccei}},
  \bibinfo{journal}{Lect. Notes Phys.} \textbf{\bibinfo{volume}{741}},
  \bibinfo{pages}{3} (\bibinfo{year}{2008}), \eprint{hep-ph/0607268}.

\bibitem[{\citenamefont{Weinberg}(1978)}]{Weinberg:1977ma}
\bibinfo{author}{\bibfnamefont{S.}~\bibnamefont{Weinberg}},
  \bibinfo{journal}{Phys. Rev. Lett.} \textbf{\bibinfo{volume}{40}},
  \bibinfo{pages}{223} (\bibinfo{year}{1978}).

\bibitem[{\citenamefont{Wilczek}(1978)}]{Wilczek:1977pj}
\bibinfo{author}{\bibfnamefont{F.}~\bibnamefont{Wilczek}},
  \bibinfo{journal}{Phys. Rev. Lett.} \textbf{\bibinfo{volume}{40}},
  \bibinfo{pages}{279} (\bibinfo{year}{1978}).

\bibitem[{\citenamefont{Jaeckel and Ringwald}(2010)}]{Jaeckel:2010ni}
\bibinfo{author}{\bibfnamefont{J.}~\bibnamefont{Jaeckel}} \bibnamefont{and}
  \bibinfo{author}{\bibfnamefont{A.}~\bibnamefont{Ringwald}},
  \bibinfo{journal}{Ann. Rev. Nucl. Part. Sci.} \textbf{\bibinfo{volume}{60}},
  \bibinfo{pages}{405} (\bibinfo{year}{2010}), \eprint{1002.0329}.

\bibitem[{\citenamefont{Svrcek and Witten}(2006)}]{Svrcek:2006yi}
\bibinfo{author}{\bibfnamefont{P.}~\bibnamefont{Svrcek}} \bibnamefont{and}
  \bibinfo{author}{\bibfnamefont{E.}~\bibnamefont{Witten}},
  \bibinfo{journal}{JHEP} \textbf{\bibinfo{volume}{06}}, \bibinfo{pages}{051}
  (\bibinfo{year}{2006}), \eprint{hep-th/0605206}.

\bibitem[{\citenamefont{Arvanitaki et~al.}(2010)\citenamefont{Arvanitaki,
  Dimopoulos, Dubovsky, Kaloper, and March-Russell}}]{Arvanitaki:2009fg}
\bibinfo{author}{\bibfnamefont{A.}~\bibnamefont{Arvanitaki}},
  \bibinfo{author}{\bibfnamefont{S.}~\bibnamefont{Dimopoulos}},
  \bibinfo{author}{\bibfnamefont{S.}~\bibnamefont{Dubovsky}},
  \bibinfo{author}{\bibfnamefont{N.}~\bibnamefont{Kaloper}}, \bibnamefont{and}
  \bibinfo{author}{\bibfnamefont{J.}~\bibnamefont{March-Russell}},
  \bibinfo{journal}{Phys. Rev. D} \textbf{\bibinfo{volume}{81}},
  \bibinfo{pages}{123530} (\bibinfo{year}{2010}), \eprint{0905.4720}.

\bibitem[{\citenamefont{Marsh}(2016)}]{Marsh:2015xka}
\bibinfo{author}{\bibfnamefont{D.~J.~E.} \bibnamefont{Marsh}},
  \bibinfo{journal}{Phys. Rept.} \textbf{\bibinfo{volume}{643}},
  \bibinfo{pages}{1} (\bibinfo{year}{2016}), \eprint{1510.07633}.

\bibitem[{\citenamefont{Raffelt and Stodolsky}(1988)}]{Raffelt:1987im}
\bibinfo{author}{\bibfnamefont{G.}~\bibnamefont{Raffelt}} \bibnamefont{and}
  \bibinfo{author}{\bibfnamefont{L.}~\bibnamefont{Stodolsky}},
  \bibinfo{journal}{Phys. Rev. D} \textbf{\bibinfo{volume}{37}},
  \bibinfo{pages}{1237} (\bibinfo{year}{1988}).

\bibitem[{\citenamefont{Atwood et~al.}(2009)}]{Fermi-LAT:2009ihh}
\bibinfo{author}{\bibfnamefont{W.~B.} \bibnamefont{Atwood}}
  \bibnamefont{et~al.} (\bibinfo{collaboration}{Fermi-LAT}),
  \bibinfo{journal}{Astrophys. J.} \textbf{\bibinfo{volume}{697}},
  \bibinfo{pages}{1071} (\bibinfo{year}{2009}), \eprint{0902.1089}.

\bibitem[{\citenamefont{Aleksi\'c et~al.}(2016)}]{MAGIC:2014zas}
\bibinfo{author}{\bibfnamefont{J.}~\bibnamefont{Aleksi\'c}}
  \bibnamefont{et~al.} (\bibinfo{collaboration}{MAGIC}),
  \bibinfo{journal}{Astropart. Phys.} \textbf{\bibinfo{volume}{72}},
  \bibinfo{pages}{76} (\bibinfo{year}{2016}), \eprint{1409.5594}.

\bibitem[{\citenamefont{Albert et~al.}(2020)}]{HAWC:2020hrt}
\bibinfo{author}{\bibfnamefont{A.}~\bibnamefont{Albert}} \bibnamefont{et~al.}
  (\bibinfo{collaboration}{HAWC}), \bibinfo{journal}{Astrophys. J.}
  \textbf{\bibinfo{volume}{905}}, \bibinfo{pages}{76} (\bibinfo{year}{2020}),
  \eprint{2007.08582}.

\bibitem[{\citenamefont{Addazi et~al.}(2022)}]{LHAASO:2019qtb}
\bibinfo{author}{\bibfnamefont{A.}~\bibnamefont{Addazi}} \bibnamefont{et~al.}
  (\bibinfo{collaboration}{LHAASO}), \bibinfo{journal}{Chin. Phys. C}
  \textbf{\bibinfo{volume}{46}}, \bibinfo{pages}{035001}
  (\bibinfo{year}{2022}), \eprint{1905.02773}.

\bibitem[{\citenamefont{De~Angelis et~al.}(2007)\citenamefont{De~Angelis,
  Roncadelli, and Mansutti}}]{DeAngelis:2007dqd}
\bibinfo{author}{\bibfnamefont{A.}~\bibnamefont{De~Angelis}},
  \bibinfo{author}{\bibfnamefont{M.}~\bibnamefont{Roncadelli}},
  \bibnamefont{and} \bibinfo{author}{\bibfnamefont{O.}~\bibnamefont{Mansutti}},
  \bibinfo{journal}{Phys. Rev. D} \textbf{\bibinfo{volume}{76}},
  \bibinfo{pages}{121301} (\bibinfo{year}{2007}), \eprint{0707.4312}.

\bibitem[{\citenamefont{Hooper and Serpico}(2007)}]{Hooper:2007bq}
\bibinfo{author}{\bibfnamefont{D.}~\bibnamefont{Hooper}} \bibnamefont{and}
  \bibinfo{author}{\bibfnamefont{P.~D.} \bibnamefont{Serpico}},
  \bibinfo{journal}{Phys. Rev. Lett.} \textbf{\bibinfo{volume}{99}},
  \bibinfo{pages}{231102} (\bibinfo{year}{2007}), \eprint{0706.3203}.

\bibitem[{\citenamefont{Simet et~al.}(2008)\citenamefont{Simet, Hooper, and
  Serpico}}]{Simet:2007sa}
\bibinfo{author}{\bibfnamefont{M.}~\bibnamefont{Simet}},
  \bibinfo{author}{\bibfnamefont{D.}~\bibnamefont{Hooper}}, \bibnamefont{and}
  \bibinfo{author}{\bibfnamefont{P.~D.} \bibnamefont{Serpico}},
  \bibinfo{journal}{Phys. Rev. D} \textbf{\bibinfo{volume}{77}},
  \bibinfo{pages}{063001} (\bibinfo{year}{2008}), \eprint{0712.2825}.

\bibitem[{\citenamefont{Mirizzi et~al.}(2007)\citenamefont{Mirizzi, Raffelt,
  and Serpico}}]{Mirizzi:2007hr}
\bibinfo{author}{\bibfnamefont{A.}~\bibnamefont{Mirizzi}},
  \bibinfo{author}{\bibfnamefont{G.~G.} \bibnamefont{Raffelt}},
  \bibnamefont{and} \bibinfo{author}{\bibfnamefont{P.~D.}
  \bibnamefont{Serpico}}, \bibinfo{journal}{Phys. Rev. D}
  \textbf{\bibinfo{volume}{76}}, \bibinfo{pages}{023001}
  (\bibinfo{year}{2007}), \eprint{0704.3044}.

\bibitem[{\citenamefont{Belikov et~al.}(2011)\citenamefont{Belikov, Goodenough,
  and Hooper}}]{Belikov:2010ma}
\bibinfo{author}{\bibfnamefont{A.~V.} \bibnamefont{Belikov}},
  \bibinfo{author}{\bibfnamefont{L.}~\bibnamefont{Goodenough}},
  \bibnamefont{and} \bibinfo{author}{\bibfnamefont{D.}~\bibnamefont{Hooper}},
  \bibinfo{journal}{Phys. Rev. D} \textbf{\bibinfo{volume}{83}},
  \bibinfo{pages}{063005} (\bibinfo{year}{2011}), \eprint{1007.4862}.

\bibitem[{\citenamefont{De~Angelis et~al.}(2011)\citenamefont{De~Angelis,
  Galanti, and Roncadelli}}]{DeAngelis:2011id}
\bibinfo{author}{\bibfnamefont{A.}~\bibnamefont{De~Angelis}},
  \bibinfo{author}{\bibfnamefont{G.}~\bibnamefont{Galanti}}, \bibnamefont{and}
  \bibinfo{author}{\bibfnamefont{M.}~\bibnamefont{Roncadelli}},
  \bibinfo{journal}{Phys. Rev. D} \textbf{\bibinfo{volume}{84}},
  \bibinfo{pages}{105030} (\bibinfo{year}{2011}), \bibinfo{note}{[Erratum:
  Phys.Rev.D 87, 109903 (2013)]}, \eprint{1106.1132}.

\bibitem[{\citenamefont{Horns et~al.}(2012)\citenamefont{Horns, Maccione,
  Meyer, Mirizzi, Montanino, and Roncadelli}}]{Horns:2012kw}
\bibinfo{author}{\bibfnamefont{D.}~\bibnamefont{Horns}},
  \bibinfo{author}{\bibfnamefont{L.}~\bibnamefont{Maccione}},
  \bibinfo{author}{\bibfnamefont{M.}~\bibnamefont{Meyer}},
  \bibinfo{author}{\bibfnamefont{A.}~\bibnamefont{Mirizzi}},
  \bibinfo{author}{\bibfnamefont{D.}~\bibnamefont{Montanino}},
  \bibnamefont{and}
  \bibinfo{author}{\bibfnamefont{M.}~\bibnamefont{Roncadelli}},
  \bibinfo{journal}{Phys. Rev. D} \textbf{\bibinfo{volume}{86}},
  \bibinfo{pages}{075024} (\bibinfo{year}{2012}), \eprint{1207.0776}.

\bibitem[{\citenamefont{Abramowski et~al.}(2013)}]{HESS:2013udx}
\bibinfo{author}{\bibfnamefont{A.}~\bibnamefont{Abramowski}}
  \bibnamefont{et~al.} (\bibinfo{collaboration}{H.E.S.S.}),
  \bibinfo{journal}{Phys. Rev. D} \textbf{\bibinfo{volume}{88}},
  \bibinfo{pages}{102003} (\bibinfo{year}{2013}), \eprint{1311.3148}.

\bibitem[{\citenamefont{Meyer et~al.}(2013)\citenamefont{Meyer, Horns, and
  Raue}}]{Meyer:2013pny}
\bibinfo{author}{\bibfnamefont{M.}~\bibnamefont{Meyer}},
  \bibinfo{author}{\bibfnamefont{D.}~\bibnamefont{Horns}}, \bibnamefont{and}
  \bibinfo{author}{\bibfnamefont{M.}~\bibnamefont{Raue}},
  \bibinfo{journal}{Phys. Rev. D} \textbf{\bibinfo{volume}{87}},
  \bibinfo{pages}{035027} (\bibinfo{year}{2013}), \eprint{1302.1208}.

\bibitem[{\citenamefont{Tavecchio et~al.}(2015)\citenamefont{Tavecchio,
  Roncadelli, and Galanti}}]{Tavecchio:2014yoa}
\bibinfo{author}{\bibfnamefont{F.}~\bibnamefont{Tavecchio}},
  \bibinfo{author}{\bibfnamefont{M.}~\bibnamefont{Roncadelli}},
  \bibnamefont{and} \bibinfo{author}{\bibfnamefont{G.}~\bibnamefont{Galanti}},
  \bibinfo{journal}{Phys. Lett. B} \textbf{\bibinfo{volume}{744}},
  \bibinfo{pages}{375} (\bibinfo{year}{2015}), \eprint{1406.2303}.

\bibitem[{\citenamefont{Meyer et~al.}(2014)\citenamefont{Meyer, Montanino, and
  Conrad}}]{Meyer:2014epa}
\bibinfo{author}{\bibfnamefont{M.}~\bibnamefont{Meyer}},
  \bibinfo{author}{\bibfnamefont{D.}~\bibnamefont{Montanino}},
  \bibnamefont{and} \bibinfo{author}{\bibfnamefont{J.}~\bibnamefont{Conrad}},
  \bibinfo{journal}{JCAP} \textbf{\bibinfo{volume}{09}}, \bibinfo{pages}{003}
  (\bibinfo{year}{2014}), \eprint{1406.5972}.

\bibitem[{\citenamefont{Meyer and Conrad}(2014)}]{Meyer:2014gta}
\bibinfo{author}{\bibfnamefont{M.}~\bibnamefont{Meyer}} \bibnamefont{and}
  \bibinfo{author}{\bibfnamefont{J.}~\bibnamefont{Conrad}},
  \bibinfo{journal}{JCAP} \textbf{\bibinfo{volume}{12}}, \bibinfo{pages}{016}
  (\bibinfo{year}{2014}), \eprint{1410.1556}.

\bibitem[{\citenamefont{Ajello et~al.}(2016)}]{Fermi-LAT:2016nkz}
\bibinfo{author}{\bibfnamefont{M.}~\bibnamefont{Ajello}} \bibnamefont{et~al.}
  (\bibinfo{collaboration}{Fermi-LAT}), \bibinfo{journal}{Phys. Rev. Lett.}
  \textbf{\bibinfo{volume}{116}}, \bibinfo{pages}{161101}
  (\bibinfo{year}{2016}), \eprint{1603.06978}.

\bibitem[{\citenamefont{Meyer et~al.}(2017)\citenamefont{Meyer, Giannotti,
  Mirizzi, Conrad, and S\'anchez-Conde}}]{Meyer:2016wrm}
\bibinfo{author}{\bibfnamefont{M.}~\bibnamefont{Meyer}},
  \bibinfo{author}{\bibfnamefont{M.}~\bibnamefont{Giannotti}},
  \bibinfo{author}{\bibfnamefont{A.}~\bibnamefont{Mirizzi}},
  \bibinfo{author}{\bibfnamefont{J.}~\bibnamefont{Conrad}}, \bibnamefont{and}
  \bibinfo{author}{\bibfnamefont{M.~A.} \bibnamefont{S\'anchez-Conde}},
  \bibinfo{journal}{Phys. Rev. Lett.} \textbf{\bibinfo{volume}{118}},
  \bibinfo{pages}{011103} (\bibinfo{year}{2017}), \eprint{1609.02350}.

\bibitem[{\citenamefont{Berenji et~al.}(2016)\citenamefont{Berenji, Gaskins,
  and Meyer}}]{Berenji:2016jji}
\bibinfo{author}{\bibfnamefont{B.}~\bibnamefont{Berenji}},
  \bibinfo{author}{\bibfnamefont{J.}~\bibnamefont{Gaskins}}, \bibnamefont{and}
  \bibinfo{author}{\bibfnamefont{M.}~\bibnamefont{Meyer}},
  \bibinfo{journal}{Phys. Rev. D} \textbf{\bibinfo{volume}{93}},
  \bibinfo{pages}{045019} (\bibinfo{year}{2016}), \eprint{1602.00091}.

\bibitem[{\citenamefont{Galanti et~al.}(2019)\citenamefont{Galanti, Tavecchio,
  Roncadelli, and Evoli}}]{Galanti:2018upl}
\bibinfo{author}{\bibfnamefont{G.}~\bibnamefont{Galanti}},
  \bibinfo{author}{\bibfnamefont{F.}~\bibnamefont{Tavecchio}},
  \bibinfo{author}{\bibfnamefont{M.}~\bibnamefont{Roncadelli}},
  \bibnamefont{and} \bibinfo{author}{\bibfnamefont{C.}~\bibnamefont{Evoli}},
  \bibinfo{journal}{Mon. Not. Roy. Astron. Soc.}
  \textbf{\bibinfo{volume}{487}}, \bibinfo{pages}{123} (\bibinfo{year}{2019}),
  \eprint{1811.03548}.

\bibitem[{\citenamefont{Galanti and Roncadelli}(2018)}]{Galanti:2018myb}
\bibinfo{author}{\bibfnamefont{G.}~\bibnamefont{Galanti}} \bibnamefont{and}
  \bibinfo{author}{\bibfnamefont{M.}~\bibnamefont{Roncadelli}},
  \bibinfo{journal}{JHEAp} \textbf{\bibinfo{volume}{20}}, \bibinfo{pages}{1}
  (\bibinfo{year}{2018}), \eprint{1805.12055}.

\bibitem[{\citenamefont{Zhang et~al.}(2018)\citenamefont{Zhang, Liang, Li,
  Liao, Feng, Yuan, Fan, and Ren}}]{Zhang:2018wpc}
\bibinfo{author}{\bibfnamefont{C.}~\bibnamefont{Zhang}},
  \bibinfo{author}{\bibfnamefont{Y.-F.} \bibnamefont{Liang}},
  \bibinfo{author}{\bibfnamefont{S.}~\bibnamefont{Li}},
  \bibinfo{author}{\bibfnamefont{N.-H.} \bibnamefont{Liao}},
  \bibinfo{author}{\bibfnamefont{L.}~\bibnamefont{Feng}},
  \bibinfo{author}{\bibfnamefont{Q.}~\bibnamefont{Yuan}},
  \bibinfo{author}{\bibfnamefont{Y.-Z.} \bibnamefont{Fan}}, \bibnamefont{and}
  \bibinfo{author}{\bibfnamefont{Z.-Z.} \bibnamefont{Ren}},
  \bibinfo{journal}{Phys. Rev. D} \textbf{\bibinfo{volume}{97}},
  \bibinfo{pages}{063009} (\bibinfo{year}{2018}), \eprint{1802.08420}.

\bibitem[{\citenamefont{Liang et~al.}(2019)\citenamefont{Liang, Zhang, Xia,
  Feng, Yuan, and Fan}}]{Liang:2018mqm}
\bibinfo{author}{\bibfnamefont{Y.-F.} \bibnamefont{Liang}},
  \bibinfo{author}{\bibfnamefont{C.}~\bibnamefont{Zhang}},
  \bibinfo{author}{\bibfnamefont{Z.-Q.} \bibnamefont{Xia}},
  \bibinfo{author}{\bibfnamefont{L.}~\bibnamefont{Feng}},
  \bibinfo{author}{\bibfnamefont{Q.}~\bibnamefont{Yuan}}, \bibnamefont{and}
  \bibinfo{author}{\bibfnamefont{Y.-Z.} \bibnamefont{Fan}},
  \bibinfo{journal}{JCAP} \textbf{\bibinfo{volume}{06}}, \bibinfo{pages}{042}
  (\bibinfo{year}{2019}), \eprint{1804.07186}.

\bibitem[{\citenamefont{Bi et~al.}(2021)\citenamefont{Bi, Gao, Guo, Houston,
  Li, Xu, and Zhang}}]{Bi:2020ths}
\bibinfo{author}{\bibfnamefont{X.-J.} \bibnamefont{Bi}},
  \bibinfo{author}{\bibfnamefont{Y.}~\bibnamefont{Gao}},
  \bibinfo{author}{\bibfnamefont{J.}~\bibnamefont{Guo}},
  \bibinfo{author}{\bibfnamefont{N.}~\bibnamefont{Houston}},
  \bibinfo{author}{\bibfnamefont{T.}~\bibnamefont{Li}},
  \bibinfo{author}{\bibfnamefont{F.}~\bibnamefont{Xu}}, \bibnamefont{and}
  \bibinfo{author}{\bibfnamefont{X.}~\bibnamefont{Zhang}},
  \bibinfo{journal}{Phys. Rev. D} \textbf{\bibinfo{volume}{103}},
  \bibinfo{pages}{043018} (\bibinfo{year}{2021}), \eprint{2002.01796}.

\bibitem[{\citenamefont{Guo et~al.}(2021)\citenamefont{Guo, Li, Bi, Lin, and
  Yin}}]{Guo:2020kiq}
\bibinfo{author}{\bibfnamefont{J.}~\bibnamefont{Guo}},
  \bibinfo{author}{\bibfnamefont{H.-J.} \bibnamefont{Li}},
  \bibinfo{author}{\bibfnamefont{X.-J.} \bibnamefont{Bi}},
  \bibinfo{author}{\bibfnamefont{S.-J.} \bibnamefont{Lin}}, \bibnamefont{and}
  \bibinfo{author}{\bibfnamefont{P.-F.} \bibnamefont{Yin}},
  \bibinfo{journal}{Chin. Phys. C} \textbf{\bibinfo{volume}{45}},
  \bibinfo{pages}{025105} (\bibinfo{year}{2021}), \eprint{2002.07571}.

\bibitem[{\citenamefont{Li et~al.}(2022)\citenamefont{Li, Bi, and
  Yin}}]{Li:2021gxs}
\bibinfo{author}{\bibfnamefont{H.-J.} \bibnamefont{Li}},
  \bibinfo{author}{\bibfnamefont{X.-J.} \bibnamefont{Bi}}, \bibnamefont{and}
  \bibinfo{author}{\bibfnamefont{P.-F.} \bibnamefont{Yin}},
  \bibinfo{journal}{Chin. Phys. C} \textbf{\bibinfo{volume}{46}},
  \bibinfo{pages}{085105} (\bibinfo{year}{2022}), \eprint{2110.13636}.

\bibitem[{\citenamefont{Cheng et~al.}(2021)\citenamefont{Cheng, He, Liang, Lu,
  and Liang}}]{Cheng:2020bhr}
\bibinfo{author}{\bibfnamefont{J.-G.} \bibnamefont{Cheng}},
  \bibinfo{author}{\bibfnamefont{Y.-J.} \bibnamefont{He}},
  \bibinfo{author}{\bibfnamefont{Y.-F.} \bibnamefont{Liang}},
  \bibinfo{author}{\bibfnamefont{R.-J.} \bibnamefont{Lu}}, \bibnamefont{and}
  \bibinfo{author}{\bibfnamefont{E.-W.} \bibnamefont{Liang}},
  \bibinfo{journal}{Phys. Lett. B} \textbf{\bibinfo{volume}{821}},
  \bibinfo{pages}{136611} (\bibinfo{year}{2021}), \eprint{2010.12396}.

\bibitem[{\citenamefont{Liang et~al.}(2021)\citenamefont{Liang, Zhang, Cheng,
  Zeng, Fan, and Liang}}]{Liang:2020roo}
\bibinfo{author}{\bibfnamefont{Y.-F.} \bibnamefont{Liang}},
  \bibinfo{author}{\bibfnamefont{X.-F.} \bibnamefont{Zhang}},
  \bibinfo{author}{\bibfnamefont{J.-G.} \bibnamefont{Cheng}},
  \bibinfo{author}{\bibfnamefont{H.-D.} \bibnamefont{Zeng}},
  \bibinfo{author}{\bibfnamefont{Y.-Z.} \bibnamefont{Fan}}, \bibnamefont{and}
  \bibinfo{author}{\bibfnamefont{E.-W.} \bibnamefont{Liang}},
  \bibinfo{journal}{JCAP} \textbf{\bibinfo{volume}{11}}, \bibinfo{pages}{030}
  (\bibinfo{year}{2021}), \eprint{2012.15513}.

\bibitem[{\citenamefont{Xia et~al.}(2018)\citenamefont{Xia, Zhang, Liang, Feng,
  Yuan, Fan, and Wu}}]{Xia:2018xbt}
\bibinfo{author}{\bibfnamefont{Z.-Q.} \bibnamefont{Xia}},
  \bibinfo{author}{\bibfnamefont{C.}~\bibnamefont{Zhang}},
  \bibinfo{author}{\bibfnamefont{Y.-F.} \bibnamefont{Liang}},
  \bibinfo{author}{\bibfnamefont{L.}~\bibnamefont{Feng}},
  \bibinfo{author}{\bibfnamefont{Q.}~\bibnamefont{Yuan}},
  \bibinfo{author}{\bibfnamefont{Y.-Z.} \bibnamefont{Fan}}, \bibnamefont{and}
  \bibinfo{author}{\bibfnamefont{J.}~\bibnamefont{Wu}}, \bibinfo{journal}{Phys.
  Rev. D} \textbf{\bibinfo{volume}{97}}, \bibinfo{pages}{063003}
  (\bibinfo{year}{2018}), \eprint{1801.01646}.

\bibitem[{\citenamefont{Pant et~al.}(2023)\citenamefont{Pant, Sunanda,
  Moharana, and S.}}]{Pant:2022ibi}
\bibinfo{author}{\bibfnamefont{B.~P.} \bibnamefont{Pant}},
  \bibinfo{author}{\bibnamefont{Sunanda}},
  \bibinfo{author}{\bibfnamefont{R.}~\bibnamefont{Moharana}}, \bibnamefont{and}
  \bibinfo{author}{\bibfnamefont{S.}~\bibnamefont{S.}}, \bibinfo{journal}{Phys.
  Rev. D} \textbf{\bibinfo{volume}{108}}, \bibinfo{pages}{023016}
  (\bibinfo{year}{2023}), \eprint{2210.12652}.

\bibitem[{\citenamefont{Li et~al.}(2021)\citenamefont{Li, Guo, Bi, Lin, and
  Yin}}]{Li:2020pcn}
\bibinfo{author}{\bibfnamefont{H.-J.} \bibnamefont{Li}},
  \bibinfo{author}{\bibfnamefont{J.-G.} \bibnamefont{Guo}},
  \bibinfo{author}{\bibfnamefont{X.-J.} \bibnamefont{Bi}},
  \bibinfo{author}{\bibfnamefont{S.-J.} \bibnamefont{Lin}}, \bibnamefont{and}
  \bibinfo{author}{\bibfnamefont{P.-F.} \bibnamefont{Yin}},
  \bibinfo{journal}{Phys. Rev. D} \textbf{\bibinfo{volume}{103}},
  \bibinfo{pages}{083003} (\bibinfo{year}{2021}), \eprint{2008.09464}.

\bibitem[{\citenamefont{Gao et~al.}(2024{\natexlab{a}})\citenamefont{Gao, Bi,
  Guo, Lin, and Yin}}]{Gao:2023dvn}
\bibinfo{author}{\bibfnamefont{L.-Q.} \bibnamefont{Gao}},
  \bibinfo{author}{\bibfnamefont{X.-J.} \bibnamefont{Bi}},
  \bibinfo{author}{\bibfnamefont{J.-G.} \bibnamefont{Guo}},
  \bibinfo{author}{\bibfnamefont{W.}~\bibnamefont{Lin}}, \bibnamefont{and}
  \bibinfo{author}{\bibfnamefont{P.-F.} \bibnamefont{Yin}},
  \bibinfo{journal}{Phys. Rev. D} \textbf{\bibinfo{volume}{109}},
  \bibinfo{pages}{063003} (\bibinfo{year}{2024}{\natexlab{a}}),
  \eprint{2309.02166}.

\bibitem[{\citenamefont{Pant}(2024)}]{Pant:2023omy}
\bibinfo{author}{\bibfnamefont{B.~P.} \bibnamefont{Pant}},
  \bibinfo{journal}{Phys. Rev. D} \textbf{\bibinfo{volume}{109}},
  \bibinfo{pages}{023011} (\bibinfo{year}{2024}), \eprint{2310.16634}.

\bibitem[{\citenamefont{Gao et~al.}(2025)\citenamefont{Gao, Bi, Li, and
  Yin}}]{Gao:2024wpn}
\bibinfo{author}{\bibfnamefont{L.-Q.} \bibnamefont{Gao}},
  \bibinfo{author}{\bibfnamefont{X.-J.} \bibnamefont{Bi}},
  \bibinfo{author}{\bibfnamefont{J.}~\bibnamefont{Li}}, \bibnamefont{and}
  \bibinfo{author}{\bibfnamefont{P.-F.} \bibnamefont{Yin}},
  \bibinfo{journal}{JCAP} \textbf{\bibinfo{volume}{01}}, \bibinfo{pages}{031}
  (\bibinfo{year}{2025}), \eprint{2407.20118}.

\bibitem[{\citenamefont{Bu and Li}(2019)}]{Bu:2019qqg}
\bibinfo{author}{\bibfnamefont{J.}~\bibnamefont{Bu}} \bibnamefont{and}
  \bibinfo{author}{\bibfnamefont{Y.-P.} \bibnamefont{Li}},
  \bibinfo{journal}{Res. Astron. Astrophys.} \textbf{\bibinfo{volume}{19}},
  \bibinfo{pages}{154} (\bibinfo{year}{2019}), \eprint{1906.00357}.

\bibitem[{\citenamefont{Armando et~al.}(2024)\citenamefont{Armando, Panci,
  Weiss, and Ziegler}}]{Armando:2023zwz}
\bibinfo{author}{\bibfnamefont{G.}~\bibnamefont{Armando}},
  \bibinfo{author}{\bibfnamefont{P.}~\bibnamefont{Panci}},
  \bibinfo{author}{\bibfnamefont{J.}~\bibnamefont{Weiss}}, \bibnamefont{and}
  \bibinfo{author}{\bibfnamefont{R.}~\bibnamefont{Ziegler}},
  \bibinfo{journal}{Phys. Rev. D} \textbf{\bibinfo{volume}{109}},
  \bibinfo{pages}{055029} (\bibinfo{year}{2024}), \eprint{2310.05827}.

\bibitem[{\citenamefont{Aleksi\'c et~al.}(2015)}]{MAGIC:2014rit}
\bibinfo{author}{\bibfnamefont{J.}~\bibnamefont{Aleksi\'c}}
  \bibnamefont{et~al.} (\bibinfo{collaboration}{MAGIC, VERITAS}),
  \bibinfo{journal}{Astron. Astrophys.} \textbf{\bibinfo{volume}{573}},
  \bibinfo{pages}{A50} (\bibinfo{year}{2015}), \eprint{1410.6391}.

\bibitem[{\citenamefont{Abe et~al.}(2023)}]{MAGIC:2022mhv}
\bibinfo{author}{\bibfnamefont{H.}~\bibnamefont{Abe}} \bibnamefont{et~al.}
  (\bibinfo{collaboration}{MAGIC}), \bibinfo{journal}{Astrophys. J. Suppl.}
  \textbf{\bibinfo{volume}{266}}, \bibinfo{pages}{37} (\bibinfo{year}{2023}),
  \eprint{2210.02547}.

\bibitem[{\citenamefont{Davies et~al.}(2021)\citenamefont{Davies, Meyer, and
  Cotter}}]{Davies:2020uxn}
\bibinfo{author}{\bibfnamefont{J.}~\bibnamefont{Davies}},
  \bibinfo{author}{\bibfnamefont{M.}~\bibnamefont{Meyer}}, \bibnamefont{and}
  \bibinfo{author}{\bibfnamefont{G.}~\bibnamefont{Cotter}},
  \bibinfo{journal}{Phys. Rev. D} \textbf{\bibinfo{volume}{103}},
  \bibinfo{pages}{023008} (\bibinfo{year}{2021}), \eprint{2011.08123}.

\bibitem[{\citenamefont{Prior and Gourgouliatos}(2019)}]{Prior:2019tnm}
\bibinfo{author}{\bibfnamefont{C.}~\bibnamefont{Prior}} \bibnamefont{and}
  \bibinfo{author}{\bibfnamefont{K.~N.} \bibnamefont{Gourgouliatos}},
  \bibinfo{journal}{Astron. Astrophys.} \textbf{\bibinfo{volume}{622}},
  \bibinfo{pages}{A122} (\bibinfo{year}{2019}), \eprint{1901.05442}.

\bibitem[{\citenamefont{Murphy et~al.}(2013)\citenamefont{Murphy, Cawthorne,
  and Gabuzda}}]{Murphy:2013zk}
\bibinfo{author}{\bibfnamefont{E.}~\bibnamefont{Murphy}},
  \bibinfo{author}{\bibfnamefont{T.~V.} \bibnamefont{Cawthorne}},
  \bibnamefont{and} \bibinfo{author}{\bibfnamefont{D.~C.}
  \bibnamefont{Gabuzda}}, \bibinfo{journal}{Mon. Not. Roy. Astron. Soc.}
  \textbf{\bibinfo{volume}{430}}, \bibinfo{pages}{1504} (\bibinfo{year}{2013}),
  \eprint{1302.0186}.

\bibitem[{\citenamefont{Dobrynina et~al.}(2015)\citenamefont{Dobrynina,
  Kartavtsev, and Raffelt}}]{Dobrynina:2014qba}
\bibinfo{author}{\bibfnamefont{A.}~\bibnamefont{Dobrynina}},
  \bibinfo{author}{\bibfnamefont{A.}~\bibnamefont{Kartavtsev}},
  \bibnamefont{and} \bibinfo{author}{\bibfnamefont{G.}~\bibnamefont{Raffelt}},
  \bibinfo{journal}{Phys. Rev. D} \textbf{\bibinfo{volume}{91}},
  \bibinfo{pages}{083003} (\bibinfo{year}{2015}), \eprint{1412.4777}.

\bibitem[{\citenamefont{Mirizzi and Montanino}(2009)}]{Mirizzi:2009aj}
\bibinfo{author}{\bibfnamefont{A.}~\bibnamefont{Mirizzi}} \bibnamefont{and}
  \bibinfo{author}{\bibfnamefont{D.}~\bibnamefont{Montanino}},
  \bibinfo{journal}{JCAP} \textbf{\bibinfo{volume}{12}}, \bibinfo{pages}{004}
  (\bibinfo{year}{2009}), \eprint{0911.0015}.

\bibitem[{\citenamefont{Galanti and Roncadelli}(2022)}]{Galanti:2022ijh}
\bibinfo{author}{\bibfnamefont{G.}~\bibnamefont{Galanti}} \bibnamefont{and}
  \bibinfo{author}{\bibfnamefont{M.}~\bibnamefont{Roncadelli}},
  \bibinfo{journal}{Universe} \textbf{\bibinfo{volume}{8}},
  \bibinfo{pages}{253} (\bibinfo{year}{2022}), \eprint{2205.00940}.

\bibitem[{\citenamefont{Pudritz et~al.}(2012)\citenamefont{Pudritz, Hardcastle,
  and Gabuzda}}]{Pudritz:2012xj}
\bibinfo{author}{\bibfnamefont{R.~E.} \bibnamefont{Pudritz}},
  \bibinfo{author}{\bibfnamefont{M.~J.} \bibnamefont{Hardcastle}},
  \bibnamefont{and} \bibinfo{author}{\bibfnamefont{D.~C.}
  \bibnamefont{Gabuzda}}, \bibinfo{journal}{Space Sci. Rev.}
  \textbf{\bibinfo{volume}{169}}, \bibinfo{pages}{27} (\bibinfo{year}{2012}),
  \eprint{1205.2073}.

\bibitem[{\citenamefont{Begelman et~al.}(1984)\citenamefont{Begelman,
  Blandford, and Rees}}]{Begelman:1984mw}
\bibinfo{author}{\bibfnamefont{M.~C.} \bibnamefont{Begelman}},
  \bibinfo{author}{\bibfnamefont{R.~D.} \bibnamefont{Blandford}},
  \bibnamefont{and} \bibinfo{author}{\bibfnamefont{M.~J.} \bibnamefont{Rees}},
  \bibinfo{journal}{Rev. Mod. Phys.} \textbf{\bibinfo{volume}{56}},
  \bibinfo{pages}{255} (\bibinfo{year}{1984}).

\bibitem[{\citenamefont{O'Sullivan and Gabuzda}(2009)}]{OSullivan:2009dsx}
\bibinfo{author}{\bibfnamefont{S.~P.} \bibnamefont{O'Sullivan}}
  \bibnamefont{and} \bibinfo{author}{\bibfnamefont{D.~C.}
  \bibnamefont{Gabuzda}}, \bibinfo{journal}{Mon. Not. Roy. Astron. Soc.}
  \textbf{\bibinfo{volume}{400}}, \bibinfo{pages}{26} (\bibinfo{year}{2009}),
  \eprint{0907.5211}.

\bibitem[{\citenamefont{Kronberg}(1994)}]{Kronberg:1993vk}
\bibinfo{author}{\bibfnamefont{P.~P.} \bibnamefont{Kronberg}},
  \bibinfo{journal}{Rept. Prog. Phys.} \textbf{\bibinfo{volume}{57}},
  \bibinfo{pages}{325} (\bibinfo{year}{1994}).

\bibitem[{\citenamefont{Blasi et~al.}(1999)\citenamefont{Blasi, Burles, and
  Olinto}}]{Blasi:1999hu}
\bibinfo{author}{\bibfnamefont{P.}~\bibnamefont{Blasi}},
  \bibinfo{author}{\bibfnamefont{S.}~\bibnamefont{Burles}}, \bibnamefont{and}
  \bibinfo{author}{\bibfnamefont{A.~V.} \bibnamefont{Olinto}},
  \bibinfo{journal}{Astrophys. J. Lett.} \textbf{\bibinfo{volume}{514}},
  \bibinfo{pages}{L79} (\bibinfo{year}{1999}), \eprint{astro-ph/9812487}.

\bibitem[{\citenamefont{Durrer and Neronov}(2013)}]{Durrer:2013pga}
\bibinfo{author}{\bibfnamefont{R.}~\bibnamefont{Durrer}} \bibnamefont{and}
  \bibinfo{author}{\bibfnamefont{A.}~\bibnamefont{Neronov}},
  \bibinfo{journal}{Astron. Astrophys. Rev.} \textbf{\bibinfo{volume}{21}},
  \bibinfo{pages}{62} (\bibinfo{year}{2013}), \eprint{1303.7121}.

\bibitem[{\citenamefont{Franceschini et~al.}(2008)\citenamefont{Franceschini,
  Rodighiero, and Vaccari}}]{Franceschini:2008tp}
\bibinfo{author}{\bibfnamefont{A.}~\bibnamefont{Franceschini}},
  \bibinfo{author}{\bibfnamefont{G.}~\bibnamefont{Rodighiero}},
  \bibnamefont{and} \bibinfo{author}{\bibfnamefont{M.}~\bibnamefont{Vaccari}},
  \bibinfo{journal}{Astron. Astrophys.} \textbf{\bibinfo{volume}{487}},
  \bibinfo{pages}{837} (\bibinfo{year}{2008}), \eprint{0805.1841}.

\bibitem[{\citenamefont{Unger and Farrar}(2024)}]{Unger:2023lob}
\bibinfo{author}{\bibfnamefont{M.}~\bibnamefont{Unger}} \bibnamefont{and}
  \bibinfo{author}{\bibfnamefont{G.~R.} \bibnamefont{Farrar}},
  \bibinfo{journal}{Astrophys. J.} \textbf{\bibinfo{volume}{970}},
  \bibinfo{pages}{95} (\bibinfo{year}{2024}), \eprint{2311.12120}.

\bibitem[{\citenamefont{Yao et~al.}(2017)\citenamefont{Yao, Manchester, and
  Wang}}]{Yao:2017kcp}
\bibinfo{author}{\bibfnamefont{J.~M.} \bibnamefont{Yao}},
  \bibinfo{author}{\bibfnamefont{R.~N.} \bibnamefont{Manchester}},
  \bibnamefont{and} \bibinfo{author}{\bibfnamefont{N.}~\bibnamefont{Wang}},
  \bibinfo{journal}{Astrophys. J.} \textbf{\bibinfo{volume}{835}},
  \bibinfo{pages}{29} (\bibinfo{year}{2017}).

\bibitem[{\citenamefont{Jansson and Farrar}(2012)}]{Jansson:2012rt}
\bibinfo{author}{\bibfnamefont{R.}~\bibnamefont{Jansson}} \bibnamefont{and}
  \bibinfo{author}{\bibfnamefont{G.~R.} \bibnamefont{Farrar}},
  \bibinfo{journal}{Astrophys. J. Lett.} \textbf{\bibinfo{volume}{761}},
  \bibinfo{pages}{L11} (\bibinfo{year}{2012}), \eprint{1210.7820}.

\bibitem[{\citenamefont{Meyer et~al.}(2021)\citenamefont{Meyer, Davies, and
  Kuhlmann}}]{Meyer:2021pbp}
\bibinfo{author}{\bibfnamefont{M.}~\bibnamefont{Meyer}},
  \bibinfo{author}{\bibfnamefont{J.}~\bibnamefont{Davies}}, \bibnamefont{and}
  \bibinfo{author}{\bibfnamefont{J.}~\bibnamefont{Kuhlmann}},
  \bibinfo{journal}{PoS} \textbf{\bibinfo{volume}{ICRC2021}},
  \bibinfo{pages}{557} (\bibinfo{year}{2021}), \eprint{2108.02061}.

\bibitem[{\citenamefont{Galanti et~al.}(2023)\citenamefont{Galanti, Roncadelli,
  and Tavecchio}}]{Galanti:2023uam}
\bibinfo{author}{\bibfnamefont{G.}~\bibnamefont{Galanti}},
  \bibinfo{author}{\bibfnamefont{M.}~\bibnamefont{Roncadelli}},
  \bibnamefont{and}
  \bibinfo{author}{\bibfnamefont{F.}~\bibnamefont{Tavecchio}},
  \bibinfo{journal}{Phys. Rev. D} \textbf{\bibinfo{volume}{108}},
  \bibinfo{pages}{083017} (\bibinfo{year}{2023}), \eprint{2301.08204}.

\bibitem[{\citenamefont{Zabalza}(2016)}]{Zabalza:2015bsa}
\bibinfo{author}{\bibfnamefont{V.}~\bibnamefont{Zabalza}},
  \bibinfo{journal}{PoS} \textbf{\bibinfo{volume}{ICRC2015}},
  \bibinfo{pages}{922} (\bibinfo{year}{2016}), \eprint{1509.03319}.

\bibitem[{\citenamefont{Tramacere et~al.}(2011)\citenamefont{Tramacere,
  Massaro, and Taylor}}]{Tramacere:2011qw}
\bibinfo{author}{\bibfnamefont{A.}~\bibnamefont{Tramacere}},
  \bibinfo{author}{\bibfnamefont{E.}~\bibnamefont{Massaro}}, \bibnamefont{and}
  \bibinfo{author}{\bibfnamefont{A.~M.} \bibnamefont{Taylor}},
  \bibinfo{journal}{Astrophys. J.} \textbf{\bibinfo{volume}{739}},
  \bibinfo{pages}{66} (\bibinfo{year}{2011}), \eprint{1107.1879}.

\bibitem[{\citenamefont{Nigro et~al.}(2022)\citenamefont{Nigro, Sitarek,
  Gliwny, Sanchez, Tramacere, and Craig}}]{Nigro:2021pxy}
\bibinfo{author}{\bibfnamefont{C.}~\bibnamefont{Nigro}},
  \bibinfo{author}{\bibfnamefont{J.}~\bibnamefont{Sitarek}},
  \bibinfo{author}{\bibfnamefont{P.}~\bibnamefont{Gliwny}},
  \bibinfo{author}{\bibfnamefont{D.}~\bibnamefont{Sanchez}},
  \bibinfo{author}{\bibfnamefont{A.}~\bibnamefont{Tramacere}},
  \bibnamefont{and} \bibinfo{author}{\bibfnamefont{M.}~\bibnamefont{Craig}},
  \bibinfo{journal}{Astron. Astrophys.} \textbf{\bibinfo{volume}{660}},
  \bibinfo{pages}{A18} (\bibinfo{year}{2022}), \eprint{2112.14573}.

\bibitem[{\citenamefont{Freeman et~al.}(2001)\citenamefont{Freeman, Doe, and
  Siemiginowska}}]{Freeman:2001uc}
\bibinfo{author}{\bibfnamefont{P.~E.} \bibnamefont{Freeman}},
  \bibinfo{author}{\bibfnamefont{S.}~\bibnamefont{Doe}}, \bibnamefont{and}
  \bibinfo{author}{\bibfnamefont{A.}~\bibnamefont{Siemiginowska}},
  \bibinfo{journal}{Proc. SPIE Int. Soc. Opt. Eng.}
  \textbf{\bibinfo{volume}{4477}}, \bibinfo{pages}{76} (\bibinfo{year}{2001}),
  \eprint{astro-ph/0108426}.

\bibitem[{\citenamefont{Wilks}(1938)}]{wilk}
\bibinfo{author}{\bibfnamefont{S.~S.} \bibnamefont{Wilks}},
  \bibinfo{journal}{The Annals of Mathematical Statistics}
  \textbf{\bibinfo{volume}{9}}, \bibinfo{pages}{60 } (\bibinfo{year}{1938}),
  \urlprefix\url{https://doi.org/10.1214/aoms/1177732360}.

\bibitem[{\citenamefont{Junk}(1999)}]{Junk:1999kv}
\bibinfo{author}{\bibfnamefont{T.}~\bibnamefont{Junk}}, \bibinfo{journal}{Nucl.
  Instrum. Meth. A} \textbf{\bibinfo{volume}{434}}, \bibinfo{pages}{435}
  (\bibinfo{year}{1999}), \eprint{hep-ex/9902006}.

\bibitem[{\citenamefont{Read}(2002)}]{Read:2002hq_cls}
\bibinfo{author}{\bibfnamefont{A.~L.} \bibnamefont{Read}}, \bibinfo{journal}{J.
  Phys. G} \textbf{\bibinfo{volume}{28}}, \bibinfo{pages}{2693}
  (\bibinfo{year}{2002}).

\bibitem[{\citenamefont{Lista}(2017)}]{Lista:2016chp}
\bibinfo{author}{\bibfnamefont{L.}~\bibnamefont{Lista}}, in
  \emph{\bibinfo{booktitle}{{2016 European School of High-Energy Physics}}}
  (\bibinfo{year}{2017}), pp. \bibinfo{pages}{213--258}, \eprint{1609.04150}.

\bibitem[{\citenamefont{Li et~al.}(2024)\citenamefont{Li, Bi, Gao, Huang, Yao,
  and Yin}}]{Li:2024ivs}
\bibinfo{author}{\bibfnamefont{J.}~\bibnamefont{Li}},
  \bibinfo{author}{\bibfnamefont{X.-J.} \bibnamefont{Bi}},
  \bibinfo{author}{\bibfnamefont{L.-Q.} \bibnamefont{Gao}},
  \bibinfo{author}{\bibfnamefont{X.}~\bibnamefont{Huang}},
  \bibinfo{author}{\bibfnamefont{R.-M.} \bibnamefont{Yao}}, \bibnamefont{and}
  \bibinfo{author}{\bibfnamefont{P.-F.} \bibnamefont{Yin}},
  \bibinfo{journal}{Chin. Phys. C} \textbf{\bibinfo{volume}{48}},
  \bibinfo{pages}{065107} (\bibinfo{year}{2024}), \eprint{2401.01829}.

\bibitem[{\citenamefont{Gao et~al.}(2024{\natexlab{b}})\citenamefont{Gao, Bi,
  Li, Yao, and Yin}}]{Gao:2023und}
\bibinfo{author}{\bibfnamefont{L.-Q.} \bibnamefont{Gao}},
  \bibinfo{author}{\bibfnamefont{X.-J.} \bibnamefont{Bi}},
  \bibinfo{author}{\bibfnamefont{J.}~\bibnamefont{Li}},
  \bibinfo{author}{\bibfnamefont{R.-M.} \bibnamefont{Yao}}, \bibnamefont{and}
  \bibinfo{author}{\bibfnamefont{P.-F.} \bibnamefont{Yin}},
  \bibinfo{journal}{JCAP} \textbf{\bibinfo{volume}{01}}, \bibinfo{pages}{026}
  (\bibinfo{year}{2024}{\natexlab{b}}), \eprint{2310.11391}.

\bibitem[{\citenamefont{Anastassopoulos et~al.}(2017)}]{CAST:2017uph}
\bibinfo{author}{\bibfnamefont{V.}~\bibnamefont{Anastassopoulos}}
  \bibnamefont{et~al.} (\bibinfo{collaboration}{CAST}),
  \bibinfo{journal}{Nature Phys.} \textbf{\bibinfo{volume}{13}},
  \bibinfo{pages}{584} (\bibinfo{year}{2017}), \eprint{1705.02290}.

\bibitem[{\citenamefont{Akaike}(1974)}]{Akaike:1974vps}
\bibinfo{author}{\bibfnamefont{H.}~\bibnamefont{Akaike}},
  \bibinfo{journal}{IEEE Trans. Automatic Control}
  \textbf{\bibinfo{volume}{19}}, \bibinfo{pages}{716} (\bibinfo{year}{1974}).

\bibitem[{\citenamefont{O'Hare}(2020)}]{AxionLimits}
\bibinfo{author}{\bibfnamefont{C.}~\bibnamefont{O'Hare}},
  \emph{\bibinfo{title}{cajohare/axionlimits: Axionlimits}},
  \bibinfo{howpublished}{\url{https://cajohare.github.io/AxionLimits/}}
  (\bibinfo{year}{2020}).

\bibitem[{\citenamefont{Acharya et~al.}(2013)}]{CTAConsortium:2013ofs}
\bibinfo{author}{\bibfnamefont{B.~S.} \bibnamefont{Acharya}}
  \bibnamefont{et~al.} (\bibinfo{collaboration}{CTA Consortium}),
  \bibinfo{journal}{Astropart. Phys.} \textbf{\bibinfo{volume}{43}},
  \bibinfo{pages}{3} (\bibinfo{year}{2013}).

\end{thebibliography}

\end{document}